\documentclass{CSML}

\def\dOi{13(1:17)2017}
\lmcsheading%
{\dOi}
{1--30}
{}
{}
{Dec.~13, 2014}
{Mar.~29, 2017}
{}



\usepackage[english]{babel}
\usepackage{hyperref,breakurl}
\usepackage{appendix}		
\usepackage{listings}
\usepackage[usenames,dvipsnames]{color}
\usepackage{graphicx, subfig}
	\DeclareGraphicsExtensions{.pdf,.png,.jpg,.jpeg,.mps,.emf}
\usepackage{epstopdf}
\usepackage{multirow}
\usepackage{xspace}	
\usepackage{soul}
\usepackage{mdwlist}
\usepackage{prooftree}
\usepackage{amsmath, amssymb,xcolor,wasysym, altbeginend}
\usepackage{comment}		
\usepackage{mathptmx,latexsym}
\usepackage{stmaryrd}
\usepackage{eurosym}
\usepackage{tabularx}
\usepackage{wrapfig}
\usepackage{breakurl}
\usepackage{enumitem}
\usepackage{calc}
\usepackage[textsize=tiny, bordercolor=black, backgroundcolor=orange, colorinlistoftodos]{todonotes}
\setlist{nolistsep}
\usepackage{multicol}
\usepackage{multirow}
\usepackage{makecell}

\usepackage{xcolor}
\definecolor{shadecolor}{RGB}{150,150,150}

\hypersetup{
  pdftitle={Multiparty Session Actors},
  pdfauthor={Rumyana Neykova and Nobuko Yoshida},
  pdfsubject={Multiparty Session Actors},
  pdfkeywords={Python, session types, Scribble, actors}
}

\begin{document}




\newcommand{\M}{\mathsf{M}}
\newcommand{\Mp}{\mathsf{M}'}

\newcommand{\monitor}{\M}

\newcommand{\A}{\mathcal{M}^{\circ}}
\newcommand{\Ap}{\mathcal{M}_2^{\circ}}


\newcommand{\Moff}[1]{\monitor^{\circ}[#1]}
\newcommand{\Moffone}[1]{\monitor_1^{\circ}[#1]}
\newcommand{\Mofftwo}[1]{\monitor_2^{\circ}[#1]}
\newcommand{\Moffp}[1]{\monitor_2^{\circ}[#1]}
\newcommand{\Mon}[1]{\monitor[#1]}
\newcommand{\Monp}[1]{\monitor'[#1]}
\newcommand{\Monone}[1]{\monitor_1[#1]}
\newcommand{\Montwo}[1]{\monitor_2[#1]}

\newcommand{\ptp}[1]{{\participant{#1}}}

\newcommand{\pf}{\__{T}}

\newcommand{\Recursion}[4]{\mu \keyword{t}.#3}
\newcommand{\RecursionT}[3]{\mu \keyword{t}\ENCan{#1}(#2).#3}
\newcommand{\RecursionP}[3]{\RecursionPAbs{#1}{#2}}
\newcommand{\RecursionPAbs}[2]{\mu X.#2}
\newcommand{\CallT}[1]{\keyword{t}}
\newcommand{\Call}[1]{\keyword{t}}
\newcommand{\CallP}[1]{X}
\newcommand{\typeconst}[1]{\mathsf{nat}}
\newcommand{\nat}{\typeconst{nat}}
\newcommand{\intTp}{\textsf{Int}}
\newcommand{\bool}{\textsf{Bool}}

\newcommand{\NTRANS}[1]{\stackrel{#1}{\not\longrightarrow}}
\newcommand{\const}[1]{\mathtt{#1}}
\newcommand{\ENTAILS}{\supset}
\newcommand{\MPSA}{{\it MPSA}}
\newcommand{\MPST}{{\it MPST}}

\newcommand{\MPSTs}{{\it MPST}s}
\newcommand{\MPSAs}{{\it MPSA}s}

\newcommand{\eveproves}{\proves_{\text{g}}}

\newif\ifny\nytrue
\nytrue
\newcommand{\NY}[1]
	{\ifny{\color{violet}{#1}}\else{#1}\fi}
\newcommand{\KH}[1]
	{\ifny{\color{red}{#1}}\else{#1}\fi}
\newcommand{\TZU}[1]
	{\ifny{\color{blue}{#1}}\else{#1}\fi}
\definecolor{dkgreen}{rgb}{0,0.5,0}
\newcommand{\MD}[1]
	{\ifny{\color{dkgreen}{#1}}\else{#1}\fi}
\newcommand{\RH}[1]
	{\ifny{\color{orange}{#1}}\else{#1}\fi}

\newcommand{\GC}{\mathcal{G}}
\newcommand{\LC}{\mathcal{T}}

\newcommand{\grmor}{\ \mathrel{\text{\large$\mid$}}\ }


\newcommand{\MPSAenv}{\envT{\Gamma}{\Delta}{D}}
\newcommand{\SPECenv}{\Theta}

\newcommand{\envTriple}[3]{#1;\, #2 #3}
\newcommand{\envT}[3]{\envTriple{#1}{#2}{#3}}
\newcommand{\envTT}[3]{\ENCan{\envTriple{#1}{#2}{#3}}}

\newcommand{\envQuintuple}[5]{#1;\, #2;\, #3;\,#4;\, #5}
\newcommand{\envQ}[5]{\envQuintuple{#1}{#2}{#3}{#4}{#5}}
\newcommand{\envQQ}[5]{\ENCan{\envTriple{#1}{#2}{#3}{#4}{#5}}}


\newcommand{\LNW}[4]{(\nu #1)(#2;\ \GQ{#3}{#4})}
\newcommand{\LNWn}[3]{(\nu \VEC{n})(#1;\ \GQ{#2}{#3})}

\newcommand{\ULNW}{\kw{N}}

\newcommand{\ULspec}{\kw{\Theta}}

\newcommand{\monitoredP}[3]{#1[#2]_{#3}}

\newcommand{\nubf}{{\pmb \nu}}

\newcommand{\MRED}{\rightarrow\!\!\!\!\!\rightarrow}

\newcommand{\AssEnv}{\ENCan{\Gamma;\Delta}}

\newcommand{\static}[1]{#1}
\newcommand{\dynamic}[1]{#1}

\newcommand{\sta}[1]{\static{#1}}
\newcommand{\dyn}[1]{\dynamic{#1}}

\newcommand{\codom}[1]{\mathsf{codom}(#1)}

\newcommand{\SHS}{Checkability}
\newcommand{\sHS}{checkability}
\newcommand{\STS}{TS}
\newcommand{\WTS}{WTS}

\newcommand{\TSWI}[3]{\STS(#1, #2, #3)}

\newcommand{\TSG}[2]{\WTS(#1)_{#2}}
\newcommand{\TSGI}[3]{\WTS(#1)_{#2}^{#3}}
\newcommand{\HSEnv}{{E}}
\newcommand{\intVar}[1]{\#(#1)}
\newcommand{\stVar}[1]{sv(#1)}
\newcommand{\stateVar}[1]{\mathtt{split_{stateVar}}(#1)}
\newcommand{\inv}[1]{\mathtt{inv}(#1)}
\newcommand{\svar}[1]{\mathtt{svar}(#1)}
\newcommand{\MYSTATEVAR}[2]{\kw{#2}}
\newcommand{\dec}{\mathbf{\mathtt{d}}}

\newcommand{\romain}[1]{$\mathbf{RD}$: #1 $\bigstar$}
\newcommand{\many}[1]{\overtilde{#1}}
\newcommand{\II}[1]{\mathtt{I}(#1)}
\newcommand{\OO}[1]{\mathtt{O}(#1)}


\newcommand{\SERVERROLE}{\texttt{S}}
\newcommand{\CLIENTROLE}{\texttt{C}}
\newcommand{\rumi}[1]{$\mathbf{RN}$ {\color{red} #1} }

\definecolor{dkblue}{rgb}{0,0.1,0.5}
\definecolor{dkgreen}{rgb}{0,0.4,0}
\definecolor{dkred}{rgb}{0.4,0,0}

\newcommand{\fgr}{Fig.}
\newcommand{\Sec}{\S}
\newcommand{\REF}[1]{\S\,\ref{#1}}

\newcommand{\MYPARAGRAPH}[1]{\paragraph{\textnormal{\textbf{#1}}}}

\definecolor{dkblue}{rgb}{0,0.1,0.5}
\definecolor{dkgreen}{rgb}{0,0.4,0}
\definecolor{dkred}{rgb}{0.4,0,0}

\newcommand{\CODESIZE}{\footnotesize}
\newcommand{\CODESTYLE}{\ttfamily}


\newcommand{\grmeq}{\; ::= \;} 
\newcommand{\Cnames}{X}
\newcommand{\cvar}{\mathtt{x}}
\newcommand{\cof}[1]{\cvar_{#1}}
\newcommand{\tconst}{c}
\newcommand{\p}{\ensuremath{\participant{p}}}
\newcommand{\q}{\ensuremath{\participant{q}}}
\newcommand{\participant}[1]{\ensuremath{\mathtt{#1}}}
\newcommand{\G}{\ensuremath{G}}

\newcommand{\SCRMATHFONT}{\fontencoding{OT1}\fontfamily{cmtt}\fontseries{m}\fontshape{n}\selectfont}

\makeatletter
\newenvironment{btHighlight}[1][]
{\begingroup\tikzset{bt@Highlight@par/.style={#1}}\begin{lrbox}{\@tempboxa}}
{\end{lrbox}\bt@HL@box[bt@Highlight@par]{\@tempboxa}\endgroup}

\newcommand\btHL[1][]{%
  \begin{btHighlight}[#1]\bgroup\aftergroup\bt@HL@endenv%
}
\def\bt@HL@endenv{%
  \end{btHighlight}%
  \egroup
}
\newcommand{\bt@HL@box}[2][]{%
  \tikz[#1]{%
    \pgfpathrectangle{\pgfpoint{1pt}{0pt}}{\pgfpoint{\wd #2}{\ht #2}}%
    \pgfusepath{use as bounding box}%
    \node[anchor=base west, fill=orange!30,outer sep=0pt,inner xsep=1pt, inner ysep=0pt, rounded corners=3pt, minimum height=\ht\strutbox+1pt,#1]{\raisebox{1pt}{\strut}\strut\usebox{#2}};
  }%
}

\makeatother

\newcommand{\hlo}[1]{ {\sethlcolor{orange} \hl{#1}} }

\lstnewenvironment{PYTHONLISTING}
{
\lstset{
  language={Python},
  showstringspaces=false,
  tabsize=2,
  commentstyle=\color{dkgreen}\itshape,
  basicstyle=\ttfamily\CODESIZE,
  keywordstyle=\color{darkblue}\CODESIZE,
  morekeywords={self,with},
  emph={@protocol, @role, protocol,send, Protocol,create, join, recv, recv_async, scope, close},
  emphstyle = \ttfamily\selectfont\color{dkgreen}\CODESIZE,
  escapeinside={*@}{@*},
  moredelim=**[is][\btHL]{`}{`}
}
}
{
}

\lstnewenvironment{SCRIBBLELISTING}%
{
\lstset{
  showstringspaces=false,
  tabsize=2,
  numbersep=2mm, 
  comment=[l]{//},
  morecomment=[s][\color{dkgreen}]{[@}{]},
  commentstyle=\color{dkgreen},
  basicstyle= \SCRMATHFONT\CODESIZE,
  keywords = {from,to, interrupt, by, choice, at, and, 
  		rec, continue, global, protocol, local, role, par, introduces, as, interruptible, or},
  	emph={string,str,int,real, bool},
  	emphstyle=\color{dkred}, 		  
  keywordstyle= \SCRMATHFONT\color{darkblue}\CODESIZE,
  mathescape = true, 
  escapeinside={*@}{@*}, 
  moredelim=**[is][\btHL]{`}{`}, 
  showlines=true,
}
}
{
}

\newcommand{\pCODE}{\lstinline[
	language=Python, 
	basicstyle=\ttfamily\small,
  keywordstyle=\color{darkblue}\small,
  morekeywords={self,with},
  emph={@protocol, @role, protocol,send, Protocol,create, join, recv, scope, recv_async, close},
  emphstyle = \ttfamily\selectfont\color{dkgreen},
  breaklines=true,
  backgroundcolor=\color{pink}, 
  mathescape=true
  ]} 

\newcommand{\CODE}{\lstinline[
	basicstyle=\SCRMATHFONT\CODESIZE,
  keywords={from,to, interrupt, by, choice, at, and, 
  rec, continue, global, protocol, local, role, par, introduces, as, interruptible},
  	emph={string,str,int, real, bool},
  	emphstyle=\color{dkred}, 		  
  keywordstyle=\SCRMATHFONT\CODESIZE\color{darkblue},
  mathescape=true, 
  breaklines=true,
  backgroundcolor=\color{darkblue}
]}

\newcommand{\mpCODE}{\lstinline[
	language=Python, 
	basicstyle=\ttfamily\scriptsize,
  keywordstyle=\color{darkblue}\scriptsize,
  morekeywords={self,with},
  emph={@protocol, @role, protocol,send, Protocol,create, join, recv, scope, recv_async,close},
  emphstyle = \ttfamily\selectfont\color{dkgreen},
  breaklines=true,
  backgroundcolor=\color{pink}, 
  mathescape=true
  ]}

\newcommand{\sCODE}{\lstinline[
	basicstyle=\SCRMATHFONT\CODESIZE,
  keywords={from,to, interrupt, by, choice, at, and, 
  		rec, continue, global, protocol, local, role, par, introduces, as, interruptible},
  	emph={string,str,int,real, bool},
  	emphstyle=\color{dkred}, 		  
  keywordstyle=\SCRMATHFONT\CODESIZE\color{darkblue},
  mathescape=true,
  breaklines=true]}

\newcommand{\tCODE}{\lstinline[
	basicstyle=\SCRMATHFONT\CODESIZE\color{dkgreen},
  keywords={from,to, interrupt, by, choice, at, and, 
  		rec, continue, global, protocol, local, role, par, introduces, as, interruptible},
  	emph={string,str,int},
  	emphstyle=\color{dkred}, 		  
  keywordstyle=\SCRMATHFONT\CODESIZE\color{darkblue},
  mathescape=true,
  breaklines=true]}



\title[Multiparty Session Actors]{Multiparty Session Actors\rsuper*}
\titlecomment{{\lsuper*}A Preliminary version of this article appeared in COORDINATION 2014}

\author[R.~Neykova]{Rumyana Neykova}	
\address{Imperial College London, 
		 Department of Computing, 
		 180 Queens Gate, London, UK}	
\email{\{rn710, n.yoshida\}@imperial.ac.uk}  

\author[N.~Yoshida]{Nobuko Yoshida}	
\address{\vspace{-18 pt}}	



\begin{abstract}
Actor coordination armoured with a suitable protocol description
language has been a pressing problem in the actors community. We
study the applicability of multiparty session type (MPST) protocols
for verification of actor programs. We incorporate sessions to
actors by introducing minimum additions to the model such as the
notion of actor roles and protocol mailboxes. The framework uses
Scribble, which is a protocol description language based on
multiparty session types. Our programming model supports actor-like
syntax and runtime verification mechanism guaranteeing communication
safety of the participating entities. An actor can implement
multiple roles in a similar way as an object can implement multiple
interfaces. Multiple roles allow for cooperative inter-concurrency
in a single actor. We demonstrate our framework by designing and
implementing a session actor library in Python and its runtime
verification mechanism. 
Benchmark results demonstrate that the runtime checks induce negligible overhead. 
We evaluate the applicability of our verification framework to 
specify actor interactions by 
implementing twelve examples from an actor benchmark suit. 
\end{abstract}

\maketitle

\section{Introduction}
\label{sec:intro}

The actor model \cite{AghaActors,ACTOR:JFP:97} is (re)gaining attention in the research community and in the mainstream programming languages as a promising concurrency paradigm. 
Unfortunately, 
the programming model itself does not ensure correct sequencing of interactions between different
computational processes. A study in \cite{TasharofiDJ13} points 
out that the property
of no shared space and asynchronous communication can make
implementing coordination protocols harder and providing a
language for coordinating protocols is needed.     
This reveals that the coordination of actors is a challenging problem when
implementing, and especially when scaling up an actor system.

To overcome this problem, we need to solve several shortcomings existing in 
the actor programming models. First, although actors often have
multiple states and complex policies for changing states, no
general-purpose specification language is currently in use for
describing actor protocols. Second, a clear guidance on actor
discovery and coordination of distributed actors is missing. 
This leads to adhoc
implementations and mixing the model with other paradigms which
weaken its benefits \cite{TasharofiDJ13}. 
Third, no verification mechanism (neither
static nor dynamic) is proposed to ensure correct sequencing of
interactions in a system of actors. Most actor implementations provide static typing
within a single actor (even with expressive pattern matching capabilities as surveyed in \S~\ref{sec:related}), but the communication between actors -- the
complex communication patterns that are most likely to deadlock --
are not checked.

We tackle the aforementioned challenges by studying applicability of
multyparty session types (MPST) \cite{HYC08}, a type theory for
communicating processes, to actor systems.  The
tenet of MPST safety assurance methodology is the use of a high-level,
global specification (also called protocol) for describing interactions between communicating entities. 
From a global protocol each entity is given a local session type which defines the order and the payload type of interactions. 
Programs are then written as a collection of possibly interleaved conversations, 
verified, against prescribed protocols and constraints, at runtime.
The practical incarnation of the theoretical MPST is the protocol specification language Scribble \cite{scribble}. Scribble is equipped with a verification mechanism, which makes protocol creation easier and
protocol verification sound. Declarative protocol specifications in Scribble can readily
avoid typical errors in communications programming, including type errors, 
disrespect of call orders, circular service dependencies and
deadlocks. 
We call these safety properties, ensured by MPST, {\em communication safety}.

Recent works from \cite{RVTool,RVPaper,DHHNY2015} prove the suitability of
Scribble and its tools for dynamic verification of real world complex protocols
\cite{OOI} and present a runtime verification framework that guarantees safety and session fidelity of the underlying communications. The verification mechanism is applied to a large
cyberinfrastructure. The MPST dynamic verification framework is built on
a runtime layer for protocol management and developers use MPST primitives for communication, which limits the verification methodology to a specific infrastructure. 

In this article, we take the MPST one step further. We adapt and distil the model to present a MPST verification of actors systems.
A main departure from our previous works \cite{RVPaper,DHHNY2015} is that they require
special conversation runtime, built into the application, which
restrict the programming style and the model applicability. In this
article, we prove the generality of MPST framework by showing Scribble
protocols offer a wider usage, in particular, for actor programming.   


Our programming model is grounded on three new design ideas: (1) use
Scribble protocols and their relation to finite state machines for
specification and runtime verification of actor interactions; (2) augment actor
messages and their mailboxes dynamically with protocol (role)
information; and (3) propose an algorithm based on virtual
routers (protocol mailboxes) for the dynamic discovery of actor
mailboxes within a protocol. We implement a session actor
library in Python to demonstrate the applicability of the approach.
To the best of our knowledge, 
this is the first design and
implementation of session types and their dynamic verification
toolchain in an actor
library. 

\paragraph{\bf Outline}
This article is
an expansion of the initial presentation of our MPST actor framework
and implementation of monitors in \cite{NeykovaY14}. Apart from the
expanded explanations, this article adds a new
section which illustrates representative actor use cases from 
the literature \cite{Savina,Savina_source}. We examine expressiveness of MPSTs for specifying actor interactions 
by implementing protocols of use cases in an extension of Scribble; 
and their monitoring in Python.  

The article is organised as follows: \S~\ref{sec:overview} gives a
brief overview of the the key features of our model and presents the
running example. \S~\ref{sec:language} describes the constructs of
Session Actors and highlights the main design decisions.
\S~\ref{sec:implementation} presents the implementation of the model
on concrete middleware. \S~\ref{sec:evaluation} evaluates the
overheads of the framework, compares it with a popular Scala actor library
\cite{AKKA} and shows applications. 
\S~\ref{sec:use_cases} presents and implements representative 
actor use cases in our framework.  
Finally, \S~\ref{sec:related}
discusses related work and concludes. The code for the runtime and the
monitor tool, example applications and omitted details are available
at \cite{OnlineAppendix}.



\section{Session Actors Programming Model}
\label{sec:overview}
This section explains how we incorporate the verification model of MPST with the programming model of actors. 
The top-down development methodology for MPST verification, which relies on a standalone tool for writing protocols in the Scribble language, is shown on the left part of Fig.~\ref{fig:framework}. The figure on the right illustrates the Session Actor programming model, which consists of annotated Python classes.

\subsection{Background}

\paragraph{\bf{The Actor model}} 
We assume the following actor features to determine our design
choices.  Actors are concurrent autonomous entities that exchange
messages asynchronously. 
An actor is an entity (a class in our framework) that has a mailbox
and a behaviour. Actors communicate between each other only by
exchanging messages. Upon receiving a message, the behaviour of the
actor is executed, upon which the actor can send messages
to other actors, create actors or change its internal
state. Each actor is equipped with a mailbox where messages are buffered. 
There are only three primitive operations each actor can perform:
(1) create new actors; (2) send messages to other actors whose address 
are known; and (3) become an actor with a new state and 
behaviour. All these operations are atomic. The only order inherent 
is a causal order, and the only guarantee provided 
is that messages sent by actors will eventually be processed
by their receivers. In addition, in our framework we enforce synchronisation and coordination
constraints.  

\paragraph{\bf{Scribble overview}} 
As a specification language we use Scribble ~\cite{scribble,scribble10}, which is 
a practical and human-readable
language for protocol specification, designed from the
multiparty session types theory~\cite{HYC08,BettiniCDLDY08}. It is a
language to describe application-level protocols among communicating
systems and aims to prevent 
communication mismatches and type errors during communications. 
A Scribble protocol represents an agreement on how
participating systems interact with each other by describing an
abstract structure of message exchanges between roles. Roles abstract
from the actual identity of the endpoints that participate in a protocol session.  

The Scribble tools include libraries for parsing, well-formedness checking algorithms
and several language bindings.
We use a Python implementation of the Scribble projection tool, available from \cite{HHNCDDY12}. The tool generates local Scribble specifications from global Scribble protocols. A local protocol specifies the message interactions from the viewpoint of only one of the participating entities. The Session Actor Framework retrieves the local protocols, stored in the file system or at a known location in the network, and use them for validation.

\definecolor{ao}{rgb}{0.0, 0.5, 0.0}
\definecolor{darkcerulean}{rgb}{0.03, 0.27, 0.49}
\definecolor{pigment}{rgb}{0.2, 0.2, 0.6}
\definecolor{darkblue}{rgb}{0.0, 0.0, 0.55}

\newcommand{\key}[1]{\mathtt{\color{darkblue}#1\color{black}}}
\newcommand{\trm}[1]{\mathit{\color{red}#1\color{black}}}
\newcommand{\nm}[1]{{\color{black}#1\color{black}}}
\newcommand{\tkey}[1]{\mathtt{\color{ao}#1\color{black}}}
\newcommand{\mesg}[2]{\mathtt{#1}(\mathtt{#2})}

The syntax of Scribble global protocols is given by the grammar below
where names are ranged over by $\tt pro$, and the messages are of the form
$\mesg{a}{T}$ with $\mathtt a$ being a label and $\mathtt T$ being the
constant type of the message exchanged (such as $\tt real, \tt bool$
and $\tt nat$).

{\small 
\[\begin{centering}\begin{array}{llll}
\trm{S} & \grmeq & \key{global \ protocol} \ \tt{pro} \ ( \key{role} \ \tt{A_1}, ..., \key{role} \ \tt{A_n} )\{\trm{\G}\} & \textit{specification} \\[0.3mm]
\trm{\G} & \grmeq & \mesg{a}{T} \  \key{from} \  \ptp A \ \key{to}\  \ptp B; \trm{\G}  & \textit{interaction} \\[0.3mm]
     &   \grmor     &  \key{choice} \ \key{at} \ \p \  \{\trm{\G_1}\}  \ \key{or} \ \ldots \ \key{or} \ \{\trm{\G_n}\}  & \textit{choice} \\[0.3mm]
     &   \grmor     &  \key{par} \ \{\trm{\G_1} \} \ \key{and} \ \ldots \ \key{and} \ \{\trm{\G_n} \}  & \textit{parallel} \\[0.3mm]
     &     \grmor   &  \key{rec} \ \tt{pro} \ \{ \trm{\G} \} & \textit{recursion}  \\[0.3mm]
     & \grmor & \key{continue} \ \tt{pro} & \textit{call} \\[0.3mm]
\end
{array}\end{centering}\]
}
A (global) specification $\trm{S}$ declares a protocol with name $\tt{pro}$, involving a list ($\tt{A_1}, .. \tt{A_n}$) of roles, and prescribing the behaviour in $\trm{\G}$. The other constructs are explained below:  
\begin{itemize}
\item An interaction, $\mesg{a}{T} \  \key{from} \  \ptp A \ \key{to}\  \ptp B; \trm{\G}$,
specifies that a message $\mesg{a}{T}$ should be sent from role $\ptp A$ to role $\ptp B$ and that the protocol should then continue as prescribed by the continuation $\trm{\G}$. 
\item A choice, $\key{choice} \ \key{at} \ \ptp A \ \{\trm{\G_1}\}  \ \key{or} \ \ldots \ \key{or} \ \{\trm{\G_n}\} $ specifies a branching where role $\ptp A$ chooses to engage in the interactions prescribed by one of the options $\trm{\G}$. The decision itself is an internal process to role $\ptp A$, i.e. how $\ptp A$ decides which scenario to follow is not specified by the protocol. 
\item A parallel, $ \key{par} \ \{\trm{\G_1} \} \ \key{and} \ \ldots \ \key{and} \ \{\trm{\G_n}\}$, specifies unordered sequence of interactions between the specified branches $\trm{\G_1}$ \ldots $\trm{\G_n}$. The interactions in $\trm{\G_1}$ interleave with any of the interactions in $\trm{\G_j}$ such that 
$j \in (1, n]$.
\item A recursion, $ \key{rec} \ \tt{pro} \ \{ \trm{\G} \}$, defines a scope with 
a name $\tt{pro}$ and a body $\trm{\G}$. Any call $\key{continue} \ \tt{pro}$ occurring inside $\trm{\G}$ executes another recursion instance. If $\key{continue} \ \tt{pro}$ is not in an appropriate scope than it remains idle. 
\end{itemize}

\noindent Causality in Scribble protocols is determined solely by the explicitly specified message transfers and, in particular, by the configuration of roles performing send and receive actions in message transfer sequences. 
Message transfers are ordered between each pair of roles. Thus, for any pair of roles $\ptp A$ and $\ptp B$ in a given protocol, messages sent by $\ptp A$ to $\ptp B$ will be received by $\ptp B$ in the same order that they were originally dispatched by $\ptp A$. Consider the following partial protocol: 

\begin{center}
\begin{tabular}{c}
\sCODE|m1() from A to B;| \\
\sCODE|m2() from C to A;| 
\end{tabular}
\end{center}

The example above requires ordering on \CODE{A}, but it does not specify synchronisation between \CODE{B} and \CODE{C}. Indeed, \CODE{C} can send the message \CODE{m2} at any time, but the endpoint code for \CODE{A} is allowed to read it from \CODE{A's} mailbox only after \CODE{A} has sent the message \CODE{m1} to \CODE{B}.

\begin{figure}[tp]
\centering
\includegraphics[scale=0.25]{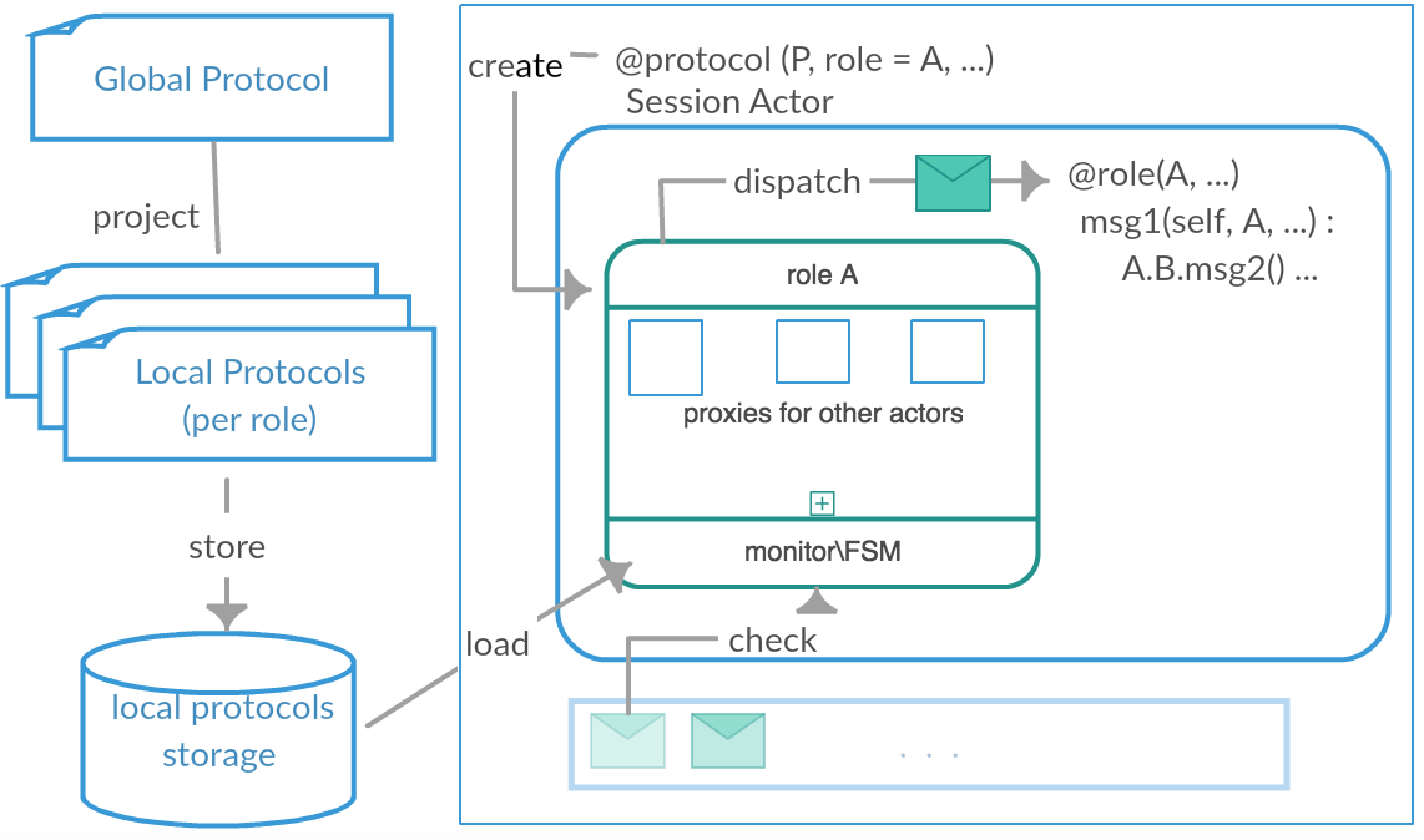}
\caption{MPST-development methodology (left) and session actors (right)}
\label{fig:framework}
\end{figure}

\paragraph{\bf{MPST verification methodology}}
The top-down development methodology for MPST verification is shown on
the left part of Fig.~\ref{fig:framework}.

A distributed protocol is specified as a global Scribble protocol, which
collectively defines the admissible communication behaviours between
the participating entities, called {\em roles}. Then,
the Scribble toolchain is used to algorithmically project the global
protocol to local specifications.

For each role a finite state
machine (FSM) is generated from the local specification. 
The FSM prescribes the allowed behaviour for each role and is used at runtime to perform validation of the allowed interactions. 
When a party requests to start or join a session, the initial message specifies
which role it implements. Its monitor retrieves the local
specification based on a protocol name and a role name. A runtime
monitor ensures that each endpoint program conforms to the core
protocol structure. 

\subsection{Session Actors Framework}
We explain how we apply the above methodology. As observed in \cite{PAM14} in many actor systems an actor encapsulates a number of passive objects, accessed from the other actors through asynchronous method calls. 
Similarly, we introduce {\em multiparty roles}, which allow multiple local control flows inside an actor. We call an actor hosting {\em multiparty roles} a Session Actor. 

\label{subsec:actormodel}
\paragraph{\bf{Session Actors}}
To verify actor interactions, each role inside an actor is associated with a FSM, generated from a global protocol. 
Roles are enabled on an actor instance via two annotations, a protocol and a role annotation. The former specifies a role declaration, while the latter specifies the role behaviour. Without the introduced annotations a session actor behaves identically to a plain actor. 

The right part of Fig.~\ref{fig:framework} shows the main actions performed by the framework and explained below:
\begin{enumerate}
   \item \CODE{create} denotes the creation of a role inside an actor. 
The \pCODE|@protocol| annotation creates a role object when the actor class is spawned. The role is associated to a protocol and contains proxies (remote references) for other actors (the other roles in the protocol) and a monitor. 
   \item \CODE{load} denotes loading a local protocol from the protocol storage when the role is created. The role monitor loads the protocol using the protocol name and generates a FSM that is used for checking. 
   \item \CODE{check} denotes checking of a message. When a message is picked up from the actor queue, it is checked by the role monitor and, if correct, dispatched to its handler.
   \item \CODE{dispatch} denotes dispatching the message to the message handler. When a message arrives in the actor queue, it is despatched to the according handler (method) according to the meta information stored in the message (a target role and a message label).
   
\end{enumerate}

\noindent An example of how the \pCODE|@role| annotation cab be used is shown below. First, consider the partial protocol
\begin{center}
\begin{tabular}{c}
\sCODE|m1() from seller to buyer;| \\
\sCODE|m2() from buyer to seller;| 
\end{tabular}
\end{center}
Then, the message transfer declaration 
\begin{center}
\sCODE|m1() from seller to buyer;| 
\end{center}
corresponds to a method declaration 
\begin{center}
\hspace{-9mm}
\pCODE|@role(seller, buyer)| \\
\pCODE|def m1(self, seller, $\ldots$)|
\end{center}

In a nutshell, an actor can be transformed to a session actor by applying the following design methodology:
\begin{enumerate}
\item the global protocol the actor is part of is written in Scribble and projected to local specifications using the Scribble commandline tool available from \cite{scribble}. The tool generates one file (local protocol) per role.
\item the actor class is annotated with a \pCODE|@protocol| decorator. 

\item each method is annotated with a \pCODE|@role| decorator. 
\item interactions with other actors are performed via the role instance.
\end{enumerate}

\paragraph{\bf{Protocol mailbox}}
Protocol mailboxes are the post offices of our systems and they exist for the duration of a protocol execution. Essentially they are virtual routers, which delegate messages according to their routing table. A protocol mailbox should be explicitly created by the developer using the \pCODE|Protocol.create| primitive. This primitive generates a fresh protocol id (the name of the virtual router) and sends the id to the actors implementing the roles for the protocol. 
The metadata of each message that is sent from a role instance (For example, \pCODE|seller.buyer.send.m2()| contains the protocol id and the target role for the message. The message is routed to the protocol mailbox matching the protocol id. The protocol mailbox delegates the message to its final recipient - the mailbox of the actor that implements the target role. 

The changes and additions that the MPST annotations bring to the original actor model 
are as follows (here we give a high level overview, while the technical details regarding their implementation are given in subsequent sections): 
\begin{enumerate}
\item[(a)] different passive objects (called roles) that run inside an actor are given a session type (\S~\ref{session_role}).
\item[(b)] an actor engages in structured protocol communications via protocol mailboxes (\S~\ref{actors_discovery}).
\item[(c)] an actor message execution is bound to a protocol structure (\S~\ref{order_checking}). This structure is checked via the internal FSM-based monitor. 
\end{enumerate}


\paragraph{\bf{Verification of Session Actors.}}


To guarantee session fidelity (actors follow the behaviour prescribed by protocols) we perform two main checks. 
First, we verify that the type (a label and a payload) of each message
matches its specification. In our implementation we have mapped message transfer labels to method invocations. For example, 
\begin{center}
\sCODE{m1(int) from A to B;} 
\end{center}
specifies that there is a method named \sCODE{m1} with an argument of type \CODE{int} in the actor, implementing role \sCODE{A}. Therefore, if such a method does not exist or its signature does not match, an error will be detected. 

Second, we verify that the overall order of
interactions is correct, i.e. interaction sequences, branches and
recursions proceed as expected, respecting the explicit dependencies.
Consider the following protocol:
\begin{center}
\sCODE{m1() from A to B;} \\
\sCODE{m2() from B to C;}
\end{center}
The protocol specifies that there is a causality between \sCODE{m1} and \sCODE{m2} at the role \sCODE{B}. More precisely, role \sCODE{B} is not allowed to send the message \sCODE{m2()} to \sCODE{C} before receiving the message \sCODE{m1()} from \sCODE{A}. 
Thus, errors, such as  communication
mismatches, that violate the permitted protocol are ruled out.

If all messages
comply to their assigned FSMs, the whole communication is guaranteed
to be safe. If an actor does not comply, violations
(such as deadlocks, communication and type mismatch)
are detected dynamically. The monitor can either block the faulty messages (outgoing faulty messages are not dispatched to the network, incoming faulty messages are not dispatched to the message handlers) or issue a warning. 
 

\subsection{Warehouse Management Use Case}
To illustrate and motivate central design decisions of our model, we
present the buyer-seller protocol from \cite{HYC08} and extend it to a
full warehouse management scenario. 
A warehouse consists of multiple customers communicating to a warehouse provider. It can be involved in a \CODE{Purchase} protocol (with customers), but can also be involved in a
\CODE{StoreLoad} protocol with dealers to update its storage. 

\paragraph{\bf{Scribble protocol.}} \ 
The interactions between the entities in the system are presented as two Scribble protocols, shown in \fgr~\ref{scribble}(left) and \fgr~\ref{scribble} (right).
The protocols are called \sCODE{Purchase} and \sCODE{StoreLoad} and involve three (a buyer~(\sCODE{B}), a seller~(\sCODE{S}) and an authenticator~(\sCODE{A})) and two (a store~(\sCODE{S}), a dealer~(\sCODE{D})) parties,
respectively.
  At the start of a purchase session, \sCODE{B}
sends login details to \sCODE{S}, which delegates the request to an
authentication server. After the authentication is completed, \sCODE{B} sends
a request quote \sCODE{req} for a product to \sCODE{S} and \sCODE{S} replies with the product
price. Then \sCODE{B} has a choice to ask for another product (\sCODE{req}), to proceed
with buying the product (\sCODE{bye}), or to quit (\sCODE{quit}). By buying a product the warehouse
decreases the product amount it has in the store. Products in stock are increased as prescribed by the \sCODE{StoreLoad} protocol,~\fgr~\ref{scribble} (right). The protocol starts with a recursion where a warehouse (in the role of \sCODE{S}) has a choice to send a product request, of the form \CODE{req(string, int)}, to a dealer \sCODE{D}. The first argument of \CODE{req} specifies the name of a requested product as a \CODE{string}, the second argument is of type \CODE{int} and stands for the number of requested products. After receiving the request, \sCODE{D} delivers the product (\sCODE{put} in line 7). These interactions are repeated in a loop until \sCODE{S} decides to \sCODE{quit} the protocol (line 10).

\paragraph{\bf Challenges.}\ 
There are several challenging points implementing the above scenario. First, a
warehouse implementation of the Store role should be involved in both protocols,
therefore it can play more than one role. Second, initially the user
does not know the exact warehouse it is buying from, therefore the
customer learns dynamically the address of the warehouse. Third, there
can be a specific restriction on the protocol that cannot be expressed
as system constants (such as specific timeout depending on the
customer). The next section explains the implementation of this example in Session Actors.

\begin{figure}[tp]
\begin{minipage}[t]{0.48\linewidth}
\lstset{numbers=left, xleftmargin=2em,framexleftmargin=1.5em}
\begin{SCRIBBLELISTING}
global protocol Purchase(
	role B, role S, role A)
{
login(str) from B to S;
login(str) from S to A;
auth(str) from A to B, S;
choice at B
	{req(str) from B to S;
	 (int) from S to B;
	}
	or
	{buy(str) from B to S
	 deliver(str) from S to B; 
	}
	or 
	{quit() from B to S;}}
\end{SCRIBBLELISTING}
\end{minipage}
\begin{minipage}[t]{0.5\linewidth}
\lstset{numbers=left, xleftmargin=2em,framexleftmargin=1.5em}
\begin{SCRIBBLELISTING}
global protocol StoreLoad
	(role D, role S)
{	
rec Rec{
	choice at S
		{req(str, int) from S to D;
	 	 put(str, int) from D to S;
continue Rec;}
	or  
	{quit() from S to D;
	 acc() from D to S;}}}	
\end{SCRIBBLELISTING}
\end{minipage}
\caption{Global protocols in Scribble for (left) Purchase protocol
and (right) StoreLoad protocol\label{scribble}}
\end{figure}

\section{Session Actor Language}
\label{sec:language}
This section explains 
the main operations in the session actor language and the 
implementation of the running example in Python. 

\begin{figure}[tp]%
\footnotesize
\begin{tabular}{ll}
Conversation API operation & Purpose \\ \hline
\mpCODE|@protocol(self_role, protocol_name,|
&  Annotating an actor class \\
\quad \quad \mpCODE|self_role, other_roles) ActorClass|
\quad\\
\mpCODE|@role(self_role, sender_role) msg_handler| &  Annotating a message handler \\
\mpCODE|Protocol.create(protocol_name, mapping)|\quad \quad 
&  Initiating a conversation \\
\mpCODE|c.role.send.method(payload)|  & Sending a message to an actor in a role\\
\mpCODE|join(self, role)| &  Actor handler for joining a protocol \\
\mpCODE|actor.send.method(payload)|  & Sending a message to a known actor\\
\end{tabular}
\caption{Session Actor operations \label{fig:api-mapping}}
\end{figure}

\paragraph{\bf Session Actor language operations.}\ 
\fgr~\ref{fig:api-mapping} presents the main session actor operations and annotations.  
The central concept of the session actor is the \textit{role}. A role
is a passive object inside an actor and it contains
meta information used for MPST-verification. A session actor is registered for a role via the \pCODE{@protocol}
annotation. The annotation specifies the name of the protocol 
and all roles that are part of the protocol. The aforementioned meta
information is stored in a role instance created in the actor. To be
used for communication, a method should be annotated with a role
using a \pCODE{@role} decorator and passing a role instance name.
The instance serves as a container for references to all protocol
roles, which allows sending a message to a role in the session actor
without knowing the actor location. A message is sent via
\pCODE{c.role.send.method}, where \pCODE{c} is the self role instance and 
\pCODE{role} is the receiver of the message.

Sending to a role without explicitly knowing the actor location is
enabled through a mechanism for actor discovery (explained in ~\Sec~\ref{actors_discovery}).  
When a session is started via the \pCODE{create}
method, actors receive an invitation for joining a
protocol. The operation \pCODE{join} is a default handler 
for invitation messages. If
a session actor changes the join behaviour and applies additional
security policies to the joiners, the method should be overloaded.
Introducing actor roles via protocol executions is a novelty of
our work: without roles and an actor discovery mechanism, actors would need
additional configurations to introduce their addresses, and this would
grow the complexity of implementations. 


\begin{figure}[tp]%
{\lstset{numbers=left, caption={Session Actor implementation for the Warehouse role},label=master_role_python, xleftmargin=0.1\textwidth, xrightmargin=0.2\textwidth}
\begin{PYTHONLISTING}
@protocol(c, Purchase, seller, buyer, auth)*@\label{l:p1}@*
@protocol(c1, StoreLoad, store, dealer)*@\label{l:p2}@*
class Warehouse(SessionActor):
	@role(c, buyer)*@\label{l:rb}@*
	def login(self, c, user): *@\label{l:l}@*
		c.auth.send.login(user)	
	
	@role(c, buyer)
	def buy(self, c, product):*@\label{l:bs}@*
		self.purchaseDB[product]-=1;
		c.buyer.send.delivery(product.details)*@\label{l:d}@*
		self.become(update, product)*@\label{l:be}@*		
	@role(c, buyer)
	def quit(self, c):
		c.buyer.send.acc()			

	@role(c1, self)		
	def update(self, c1, product):*@\label{l:us}@*
		c1.dealer.send.req(product, n)*@\label{l:ue}@*				
		
	@role(c1, dealer)
	def put(self, c1, product): *@\label{l:p}@*
		self.purchaseDB[product]+=1:			

def start_protocols():
	c = 	Protocol.create(Purchase, {seller:Warehouse, *@\label{l:cp1}@*
buyer: Customer, auth: Authenticator})
  c1 = Protocol.create(StoreLoad, {store: c.seller, *@\label{l:cp2}@*
 dealer: Shop})
	c.buyer.send.start() *@\label{l:j1}@*	
	
\end{PYTHONLISTING}}
\end{figure}

\paragraph{\bf Warehouse service implementation.}
We explain the main constructs through an implementation of a session actor for the Warehouse service. 
Listing~\ref{master_role_python} presents the
implementation of a warehouse service as a single session
actor that keeps the inventory as a state (\pCODE{self.purchaseDB}). Lines~\ref{l:p1}--\ref{l:p2} annotate the
session actor, e.g  class \pCODE{Warehouse},  with two protocol decorators -- \pCODE{c} and \pCODE{c1} (for
seller and store roles respectively). The instances \pCODE{c} and \pCODE{c1} are accessible
within the warehouse actor and are holders for mailboxes of the
other actors, involved in the two protocols. 

All message handlers are
annotated with a \pCODE{@role} and are implemented as methods.
For example, the login method (line~\ref{l:l}) is invoked when a \pCODE{login}
message (line 4,~\fgr~\ref{scribble} (left)) is sent. The role
annotation for \pCODE{c} (line~\ref{l:rb}) specifies the originator of the message to be \pCODE{buyer}. 

The handler body continues following line 5,~\fgr~\ref{scribble} (left) --
sending a \pCODE{login} message via the \pCODE{send} primitive to the session
actor, registered as a role \pCODE{auth} in the protocol of \pCODE{c}. Value \pCODE{c.auth}
is initialised with the \pCODE{auth} actor mailbox as a result of the
actor discovery mechanism (explained in~\S~\ref{actors_discovery}). The
handling of \pCODE{auth} (line 6,~\fgr~\ref{scribble} (left)) and
\pCODE{req} (line 6,~\fgr~\ref{scribble} (right)) messages is similar, so
we omit it and focus on the \pCODE{buy} handler (line~\ref{l:bs}--\ref{l:be}), where after sending
the delivery details (line~\ref{l:d}), the warehouse actor sends a message to
itself (line~\ref{l:be}) using the primitive \pCODE{become} with value \pCODE{update}.
Value \pCODE{update} is annotated with another role \pCODE{c1}, but has as a sender
\pCODE{self}. This is the mechanism used for switching between roles within
an actor. Update method (line~\ref{l:us}--\ref{l:ue}) implements the
request branch (line 6--8,~\fgr~\ref{scribble} (right)) of the
\pCODE{StoreLoad} protocol -- sending a request to the \pCODE{dealer} and handling
the reply via method \pCODE{put}.

The correct order of messages is verified by FSMs, embedded in \pCODE{c} and 
\pCODE{c1} respectively. As a result, errors, such as calling \pCODE{put} before \pCODE{update} or executing two consecutive updates, will be detected as invalid. 

A protocol is started using a function \pCODE{Protocol.create}. It takes, as parameters,  (1) a name of a protocol and (2) a mapping between role names and actor classes. Note that, for readability, we use more descriptive role names --\pCODE{buyer}, \pCODE{seller} and \pCODE{auth} instead of \CODE{B}, \CODE{S}, \CODE{A} respectively. 
Creating a protocol involves (1) creating a protocol mailbox with a newly generated protocol id; (2) spawning all actors (if they still have not been started) involved in the protocol; and (3) sending the address of the protocol mailbox (the protocol id) to all actors. Lines~\ref{l:cp1} starts the \pCODE{purchase} protocol and spawns the Warehouse, Customer and Authenticator actors (Warehouse, Customer and Authenticator are names of actor classes).
Lines~\ref{l:cp2} starts the \pCODE{storeLoad} protocol. Since the Warehouse actor is already spawned we pass a reference to an existing actor, e.g \pCODE|c.seller|.
Then, the code for starting a role, \pCODE{c.buyer.send.start()}, sends a message to \pCODE{buyer} to start the protocol interactions as the global protocol prescribes (the protocol prescribes the first interaction is from \pCODE{buyer} sending a \pCODE{login} message to \pCODE{seller}).



\section{Implementations of Session Actors}
\label{sec:implementation}
This section explains our implementation of Session Actors. 
The key design choices follow the actor framework explained 
in \S~\ref{subsec:actormodel}.
We have implemented the multiparty session actors on top of Celery
\cite{celery} (a Python framework for distributed task processing) with support for
distributed actors \cite{cell}. Celery uses advanced message queue protocol (AMQP
0-9-1 \cite{AMQP}) as a transport. The reason for choosing AMQP
network as base for our framework is that AMQP middleware shares a
similar abstraction with the actor programming model, which makes the
implementation of distributed actors more natural.

\subsection{AMQP Background}\ 
\label{amqp}
We first summarise the key features of the AMQP model. In AMQP, messages
are published by producers to entities, called exchanges (or
mailboxes). Exchanges then distribute message copies to queues using
binding rules. Then AMQP brokers (virtual routers) deliver
messages to consumers subscribed to queues.  Exchanges can be of
different types, depending on the binding rules and the routing strategies they implement.
We use three exchange types in our implementation: {\em round-robin} exchange
(deliver messages alternating  
all subscribers), {\em direct} exchange (subscribers subscribe with a binding key and
messages are delivered when a key stored in the message meta
information matches the binding key of the subscription) and {\em broadcast} exchange
(deliver a message to all subscribers).

Distributed actors are naturally represented in this AMQP context
using the abstractions of exchanges. Each actor type is represented in
the network as an exchange and is realised as a consumer subscribed to
a queue based on a pattern matching on the actor id.

%


\subsection{Actor Roles}
\label{session_role}
Our extensions to
the actor programming model are built using Python advanced
abstraction capabilities: two main capabilities are {\em greenlets} (for
realising the actors inter-concurrency) and {\em decorators} (for annotating
actor types and methods). 

A greenlet (or micro/green thread) is a
light-weight cooperatively-scheduled execution unit in Python.  A
Python decorator is any callable Python object that is used to modify
a function, method or class definition, annotated with 
\pCODE{@} symbol. A decorator is passed the original object being
defined and returns a modified object, which is then bound to the name
in the definition.  The decorators in Python are partly inspired by Java
annotations.

A key idea of actor roles is that each role runs as a micro-thread in
an actor (using Python greenlet library). A role is activated when a
message is received and ready to be processed. Switching between roles
is done via the \pCODE{become} primitive (as demonstrated in
Listing~\ref{master_role_python}), which is realised as sending a message
to the internal queue of the actor. Roles are scheduled cooperatively.
This means that at most one role can be active in a session actor at a
time.

Actors are assigned roles by adding the \pCODE{@protocol} decorator to the actor class declaration. 
The annotation 
\begin{center}
\begin{tabular}{c}
\begin{PYTHONLISTING}
@protocol(seller, "Purchase", "seller", ["buyer"])
class Warehouse
\end{PYTHONLISTING}
\end{tabular}
\end{center}
creates a private field \pCODE{seller} on the \pCODE{class Warehouse}. The field has a type \pCODE{ActorRole}. \pCODE{ActorRole} is a custom class that encapsulates an actor role behaviour. In this case, the behaviour of the \pCODE{seller} role. We recall \fgr~\ref{fig:framework}, which illustrates the structure of a Session Actor. In addition to a protocol name, and an \pCODE{ActorRole} instance (role \CODE{A} in \fgr~\ref{fig:framework}), a Session Actor also contains a FSM monitor and addresses to other actors in the protocol. Thus, the other participating actors in the protocol are accessed using role names (for example, \pCODE{seller.buyer} returns the address of the actor that implements \pCODE{buyer}). 

All communication interactions on the role instance \pCODE{seller} should follow the interactions prescribed by the protocol \pCODE{Purchase} for the role \pCODE{seller}. Therefore: 
\begin{enumerate}
\item all labels sent to the role \pCODE{seller} in the protocol specification, are implemented as methods of the class \pCODE{Warehouse} and  
\item the above methods are annotated with \pCODE{@role} decorator. 
\end{enumerate}

\noindent For example, the specification that a \pCODE{buyer} sends a message \pCODE{quit} to the \pCODE{seller} requires that there is a method \pCODE{quit} annotated with \pCODE{@role(seller, buyer)}. The latter can be read as \pCODE{@role(receiver_role=seller,  sender_role=buyer)}. 

Messages received in an actor mailbox are dispatched to the matching handlers (methods) only if the \pCODE{@role} decorator for the method corresponds to the meta information about the sender and receiver roles in the message header. For example, the message dispatch as a result of the call 
\begin{center}
\pCODE{buyer.seller.send.quit()} 
\end{center}
will invoke the body of the method declared as: 
\begin{center}
\pCODE{@role(seller, buyer) def quit}.
\end{center}

\subsection{Actors Discovery} 
\label{actors_discovery}
\begin{figure}[tp]
\begin{minipage}[b][][b]{0.4\linewidth}
\includegraphics[scale=0.4]{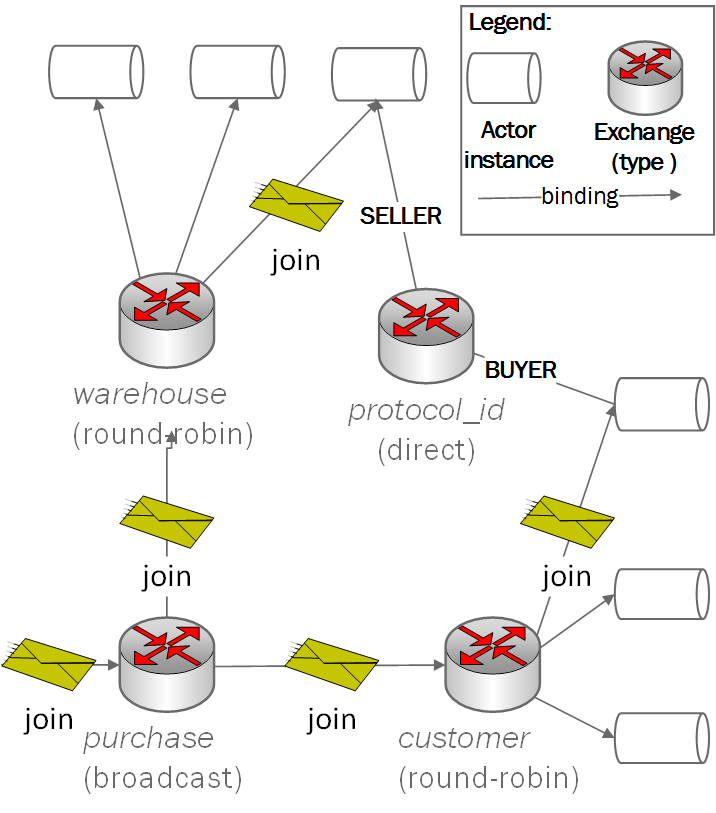}
\caption{Organising Session Actors into protocols\label{fig:actor_into_protocol}}
\label{network}
\end{minipage}
\begin{minipage}[b][][b]{0.4\linewidth}
\hspace{5mm}
  \includegraphics[scale=0.7]{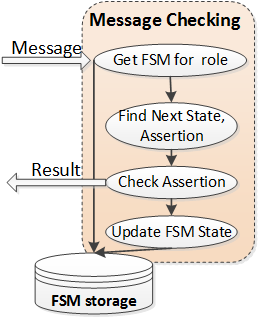}  
\vspace{15pt}
  \caption{Session Actors Monitoring \label{fig:actors_monitoring}}
\end{minipage}
\end{figure}
Fig.~\ref{network} presents the network setting (in terms of AMQP objects) for realising the actor discovery for \CODE{buyer} and \CODE{seller} of the protocol \CODE{Purchase}. We use the three types from AMQP explained in \REF{amqp}. 
For simplicity, we create some of the objects on starting of the actor system -- round-robin exchange per actor type (\CODE{warehouse} and \CODE{customer} in Fig.~\ref{network}) and broadcast exchange per protocol  (\CODE{purchase} in Fig.~\ref{network}). All spawned actors alternate to receive messages addressed to their exchange. When session actors are started their exchange is bound to the protocol exchange according to their \pCODE{@protocol} annotations (line~\ref{l:p1} in Listing~\ref{master_role_python} binds \CODE{warehouse} exchange to \CODE{purchase} exchange, as shown in Fig.~\ref{network}).

An exchange accepts messages from a producer application and routes them to message queues. Exchanges are the ``mailboxes'' of AMQP. Unlike other popular middlewares, in AMQP messages are not published directly to queues. Messages are published to exchanges that route them to queue(s) using pre-arranged criteria called bindings.

Exchanges are very cheap to create with a throughput of one million messages per second\footnote{\burl{http://www.rabbitmq.com/blog/2011/03/28/very-fast-and-scalable-topic-routing-part-2/}}. Therefore we do not consider their use as a bottleneck of our implementation. Messages are not stored in exchanges and therefore only the exchange throughput evaluation is relevant to performance concerns. 

The session initiation takes  advantage of the available AMQP infrastructure. The actor discovery wraps the boilerplate code for spawning new actors, creating new session id and exchanging actor addresses between participants. The exact exchange bindings for our running example are explained below.

We now explain the workflow for actor discovery. When a protocol is started, a fresh protocol id and an exchange with that id are created.  
An AMQP direct exchange (see \REF{amqp}) is used so that
messages with a routing key are delivered to actors linked to the
exchange with binding to that key 
 (it corresponds to \CODE{protocol_id} in \fgr~\ref{network}). 
 Then, a \CODE{join} message is sent to the protocol exchange and
 delivered to one actor per registered role (\CODE{join} is broadcasted
 to \CODE{warehouse} and \CODE{customer} in \fgr~\ref{network}). On \CODE{join}, an actor binds itself to the \CODE{protocol_id} exchange with
 subscription key equal to its role (bindings \pCODE{seller} and \pCODE{buyer} in \fgr~\ref{network}). When an actor sends a message to
 another actor within the same session (for example \pCODE{c.buyer.send.ack()}
 in Listing~\ref{master_role_python}), the message is sent to the
 protocol id exchange (stored in \pCODE{c}) and from there delivered to
 the \pCODE{buyer} actor.

\subsection{Preservation through FSM Checking} 
\label{order_checking}
Before a message is dispatched to its message handler, the message goes
through a monitor. Fig.~\ref{fig:actors_monitoring} illustrates the
monitoring process. Each message contains meta information (a
routing key) with a role name and a 
protocol id. When an actor joins a
protocol, the FSM, generated from the Scribble compiler (as shown in
Fig.~\ref{fig:framework}) is loaded from a distributed storage to the
actor memory. 

Message checking goes through the following steps. First, depending on a
role and a protocol id a matching FSM is retrieved from the actor
memory. Next, the FSM checks the message labels/operators (that are
part of the actor payload) and sender and receiver roles (that are
part of the message binding key) are
valid. 

The \textit{check assertions} step verifies that if any
constraints on the size/value of the payload are specified in
Scribble, they are also fulfilled. If a message is detected as wrong
the session actor throws a distributed exception and sends an error
message back to the sending role and does not pass the message to its
handler for processing. This behaviour can be changed by implementing the
\CODE{wrong_message} method of the session actor. 






\section{Evaluations of Session Actors}
\label{sec:evaluation}
This section reports on the performance of our framework. The goal of
our evaluation is two fold. 
First, we compare our host distributed actor framework \cite{celery}
with a mainstream actor library (AKKA \cite{AKKA}). 
Second, we show that our main
contribution, runtime verification of Scribble protocols, can be realised with
reasonable cost. The full source code of the benchmark protocols and
applications and the raw data are available from the project page
\cite{OnlineAppendix}.

\subsection{Session Actors Performance}
\label{subsec:performance}

\begin{figure}[tp]
\begin{minipage}[t!]{0.53\linewidth}
\begin{flushleft}
\hspace*{-1cm}
\includegraphics[scale=0.3]{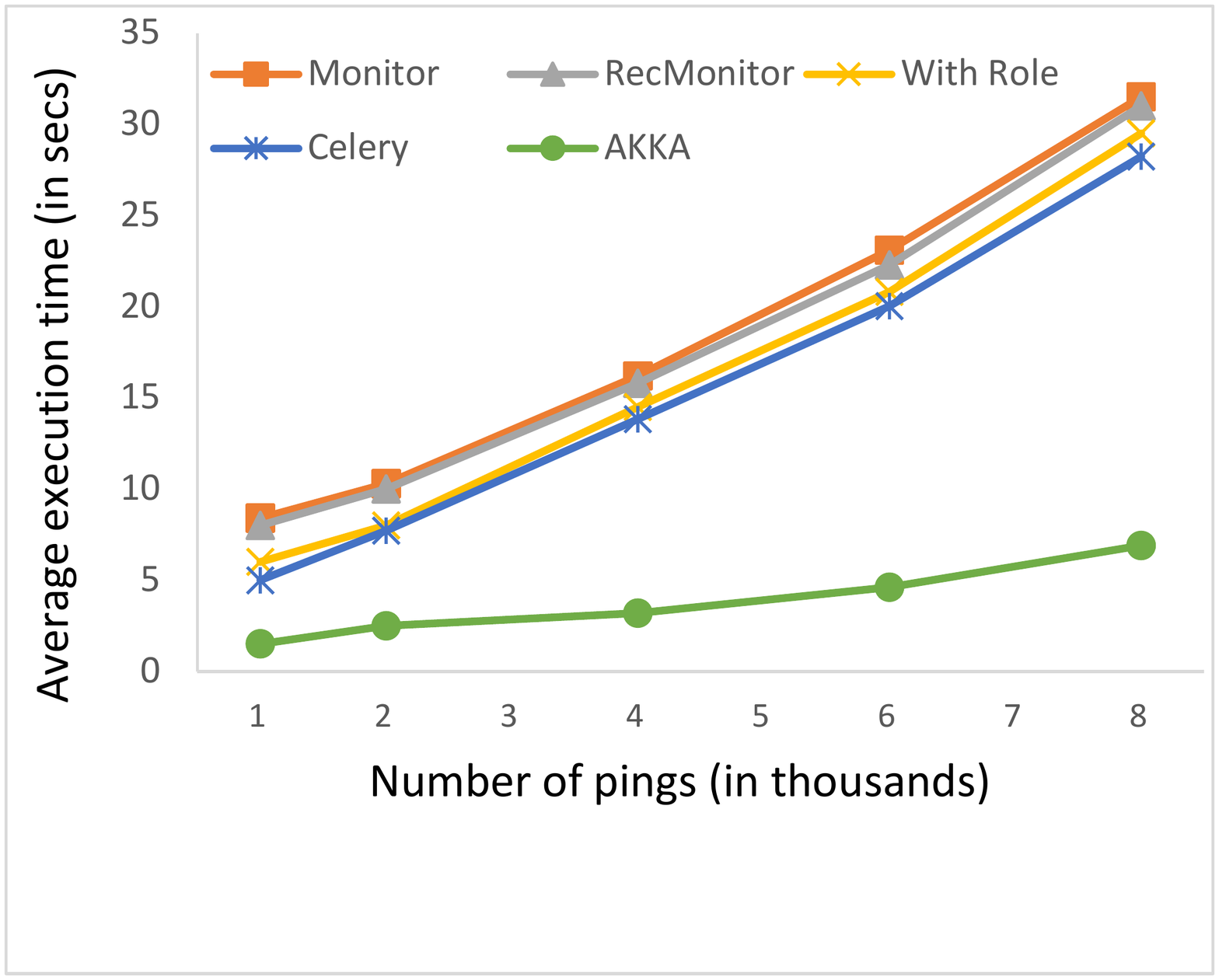}
\end{flushleft}
\end{minipage}
\begin{minipage}[t!]{0.45\linewidth}
\begin{flushright}
\begin{SCRIBBLELISTING}
global protocol FlatPingPong(
	role C, role S)
{
ping(str) from C to S;
pong(str) from S to C;
}


global protocol RecPingPong(
	role C, role S)
{
rec Loop{	
	ping(str) from C to S;
	pong(str) from S to C;
continue Loop; }
}
\end{SCRIBBLELISTING}
\end{flushright}
\end{minipage}
\caption{The PingPong benchmark, showing the overhead of message delivery (left) and The FlatPingPong and RecPingPong protocol (right)}\label{fig:benchmarks}
\end{figure}

We test the overhead of message delivery in our implementation using the
pingpong benchmark \cite{TasksActors} in which two processes send each
other messages back and forth.  The original version of the code was
obtained from Scala $\mathtt{pingpong.scala}$\footnote{\url{http://scala-lang.org/old/node/54}} and adapted to
use distributed AKKA actors (instead of local). 
We distinguish two
protocols. Each pingpong can be a separate session (protocol
FlatPingPong) or the whole iteration can be part of one recursive
protocol (RecPingPong). The protocols are given in \fgr~\ref{fig:benchmarks} (right). 
This distinction is important only to session
actors, because the protocol shape has implications on checking. For
AKKA actors the notion of session does not exist and therefore the two
protocols have the same implementation.

\paragraph{\bf{Set up.\ }}
We run each scenario 50 times and measured the overall execution
time (the difference between successive runs was found to be
negligible). The creation and population of the network was not
measured as part of the execution time. The client and server actor
nodes and the AMQP broker were each run on separate machines (Intel
Core2 Duo 2.80 GHz, 4 GB memory, 64-bit Ubuntu 11.04, kernel 2.6.38).
All hosts are interconnected through Gigabit Ethernet and latency
between each node was measured to be 0.24 ms on average (ping 64
bytes). 
The version of Python used was 2.7, of Scala -- 2.10.2 and of the AKKA-actors -- 2.2.1.

\paragraph{\bf{Results.\ }}
Fig.~\ref{fig:benchmarks} (left) compares three separate benchmark
measurements for session actors. The base case for
comparison, annotated as "Celery", is a pure actor implementation without
the addition of roles. "With Role" measures the overhead of roles
annotations without having the monitor capabilities enabled (without verifying communication safety). The two
main cases, "Rec Monitor" and "Monitor", measure the full overhead
of session actors. This separation aims to clearly illustrate the
overhead introduced by each of the additions, presented in this chapter: roles annotations, Scribble-based runtime verification and actor discovery. Note that
FSMs for the recursive and for the flat protocol have the same
number of states. Therefore, the observed difference in the
performance is a result of the cost of actor discovery.

The  difference between the performance of AKKA and
Celery is not surprising and can be explained by the distinct nature
of the frameworks. AKKA runs on top of JVM and the remote actors 
in these benchmarks use direct TCP connections. On the other hand,
Celery is a Python-based framework which uses a message middleware as a
transport, which adds a layer of indirection. Given these
differences, Celery actors have reasonable performance and
are a viable alternative.

Regarding our additions we can draw several positive conclusions from the benchmarks: (1) the cost of FSM checking is negligible and largely
overshadowed by the communication cost (such as latency and routing); 
and (2) the cost of the mechanism for actor discovery is reasonable, given the protocol load.

\subsection{Monitoring Overhead}
\label{subsec:overhead}
In this subsection, 
we discuss the row overhead of monitor checking, which is
a main factor that can affect the performance and scalability of our
implementation. Knowing the complexity of checking protocols of
different length is useful to reason about the applicability of the model.

We have applied two optimisations to the monitor, presented in \Sec~\ref{order_checking}, which significantly reduce the runtime
overhead. First, we pre-generate the FSM based on the global protocol.
Second, we cache the FSM protocols the first time when a role is loaded.
Therefore, the slight runtime overhead, observed in the previous
section is a result of the time required to choose the correct FSM
among all in-memory FSMs inside the actor and the time required to
check the FSM transition table that the message is expected. Both have linear complexity on the number of roles running inside an 
actor and the number of states inside a FSM respectively. Table~\ref{table:overhead} shows the approximate execution time for checking a protocols of increasing length of interactions. The protocol for testing is the request-reply protocol, given below:
\begin{center}
\sCODE|msg1() from A to B; msg2() from B to A; $\ldots$ msg_n() from B to A|)
\end{center}
We evaluate the above protocol for $n$ ranging over 100, 1000 and 10 000 respectively. Then we report the time taken to complete each of the protocols. We omit results for $n$ smaller than 100 since the execution time is negligible. 

\begin{table}[t]
\centering
\begin{tabular}{ |l| l l l|}
  \hline
  \quad States  \quad & \quad 100  & \quad 1000  & \quad 10000\\
  \quad Time (ms) \quad  & \quad 0.0479 \quad & \quad 0.0501  \quad  & \quad 0.0569 \quad \\
\hline
\end{tabular}
\vspace{2mm}
\caption{Execution Time for checking protocol with increasing length}
\label{table:overhead}
\end{table}

\subsection{Applications of Session Actors}
\label{subsec:applications}
As a practical evaluation of our framework, we have implemented two
popular actor applications and adapted them to session actors. 

The
first application is a distributed chat with multiple chat
rooms, where a number of clients connect to a reactive server and
execute operations based on their privileges. Depending on privileges
some clients might have extended number of allowed messages. We impose
this restriction dynamically by annotating the restricted operations
with different roles. An operation is allowed only if its actor has
already joined a session in the annotated role. 

The second use case is a \CODE{MapReduce} protocol for counting words in a document. It is an adaptation from an example presented in the official website for celery actors\footnote{\burl{http://cell.readthedocs.org/en/latest/getting-started/index.html}}. The session actor usage removes the requirement for the manual actor spawning of \CODE{Reducers} actors inside the \CODE{Mapper} actor, which reduces the code size and the complexity of the implementation.

\begin{center}
\lstset{numbers=left, caption={The WordCount protocol and its implementation in Python}, 
label=map_reduce, boxpos=c,xleftmargin=0.2\textwidth, xleftmargin=0.2\textwidth}
\begin{PYTHONLISTING}
#=======Code For The Mapper Role=========
@protocol(c, WordCount, R, A, M)
class Mapper(Actor):
	@role(c)	 # invoked on protocol start	
	def join(self, c, file, n):
		self.count_document(self,file, n)
			
	@role(c, self)	
	def count_document(self, c, file, n):
		with open(file) as f:
			lines = f.readlines(), count = 0
			for line in lines:
				reducer = c.R[count%n], count+=1
				reducer.send.count_lines(line)

#=======Code For Creating a protocol=======			
c = Protocol.create(WordCount, {M=Mapper, 
	A=Aggregator}, {R:(Reducer, 10)})
c.M.send.join(file="file1.txt", n=10)

#=======Code For Aggregator role===========			
@protocol(c, WordCount, A, M, R)
class Aggregator(Actor):
	@role(c, master)
	def aggregate(self, c, words):
	for word, n in words.iteritems():
		self.result.setdefault(word, 0)
		self.result[word] += n 
		# when a treshhold is reached 
		# print the results
	c.close()

#=======Code For Reducer role==============			
@protocol(c, WordCount, R, A, M)
class Reducer(Actor):
	@role(c, M)        
	def count_lines(self, c, line):
	words = {}
		for word in line.split(" "):
		words.setdefault(word, 0)
		words[word] += 1
		c.A.send.aggregate(words)
\end{PYTHONLISTING}
\end{center}

We give in Listing~\ref{map_reduce} the implementation of the latter example. For sake of space and clarity, the implementation reported here abstracts from technical details that are not important for the scope of this work. The protocol is started on line 17 specifying actors for each role and passing the arguments for the initial execution -- \CODE{n} (the number of reducers) and \pCODE{file} (the name of the source file to be counted). The \pCODE{Mapper} implements the \pCODE{join} method, which is invoked when an actor joins a protocol. Since the \pCODE{Mapper} is the starting role, when it joins, it sends a message to itself to proceed with the execution of \pCODE{count_document}. Note that spawning and linking of actors are not specified in the code because the actor discovery mechanism accounts for it. The protocol proceeds by \pCODE{Mapper} sending one message for each line of the document. The receiver of each message is one of the \pCODE{Reducers}. Each \pCODE{Reducer} counts the words in its line and sends the result to the \pCODE{Aggregator} (\pCODE{c.A}, line 42), which stores all words in a dictionary and aggregates them. When a threshold for the result is reached, the \pCODE{Aggregator} prints the result and stops the session explicitly. 

Our experiences with session actors are promising. Although they
introduce new notions, i.e.~a protocol and a role, we found that
their addition to a class-based actor framework is 
natural to integrate without requiring a radical new way of
thinking. The protocol and role annotations are matched well with typical
actor applications, and they even result in simplifying 
the code by removing the
boilerplate for actor discovery and coordination. 
The protocol-oriented approach to actor programming accounts for the early
error detection in the application design and the coordination of
actors. The runtime verification guarantees and enforces the correct
order of interactions, which is normally ensured only by hours of
testing.


\section{Use Cases}
\label{sec:use_cases}
In this section we evaluate the applicability of our framework 
to verify use cases which contain common communication patterns found in actor
languages.  We select the use cases from an actor
benchmark suit, Savina \cite{Savina}. Savina aims to identify a
representative set of actor applications. The use cases range from popular micro benchmarks to
classic concurrency problems. Each benchmark tests a different aspect
of the actor implementation (e.g.~actor creation, throughput, mailbox
contention) and presents a different interaction patterns
(e.g.~master-slave, round-robin communication, synchronous request-reply).

We express twelve out of the sixteen benchmarks, presented in the
micro and concurrency benchmarks sections of the suit. 
We observe that standard primitives for multiparty session
types (recursion, choice, point-to-point interaction and parallel
composition) are not sufficient to specify most of the use cases, but
more advanced features such as assertions \cite{BHTY10}, subsessions
\cite{DBLP:conf/concur/DemangeonH12} and dynamic role creation \cite{DY11} leverage
the expressive capabilities of Scribble. 
The required extensions are already 
proposed in the theoretical literature of MPSTs, and can be expressed
using combinations of various theoretical formalisms. 
Here we (1) identify 
(a combination of) the MPST extensions needed to express 
actors patterns; (2) specify 
them in Scribble; and (3) implement end-point programs in Python. 


\begin{table}\footnotesize
\begin{tabular}{|l|l|l|l|l|}
\hline
No & Use Case Name & MPST feature required & paper \\
\hline
1 & Ping Pong & standard & \cite{scribble10}\\ 
\hline
2 & Counting Actor & standard & \cite{RVTool}\\
\hline 
3 & Fibonacci & recursive subsession with dynamic role creation & \cite{DBLP:conf/concur/DemangeonH12} \\
\hline
4 & Big & parallel;  join & \cite{scribble10} \\
\hline
5 & Thread Ring & Pabble (parameterised MPST) & \cite{Pabble} \\
\hline
6 & KFork (throughput) & subsessions & \cite{DBLP:conf/concur/DemangeonH12} \\
\hline
7 & Dining Philosophers & parameterised parallel; group role & new \\
\hline
8 & Sleeping Barber & parameterised parallel; group role & new \\
\hline 
9 & Logistic Map Series & parameterised parallel with join & new \\
\hline 
10 & Bank Transaction & parallel composition with join & new \\
\hline
11 & Cigarette Smoker & parameterised parallel; group role  & new \\
\hline
\end{tabular}
\caption{List of implemented Savina benchmarks}
\label{table:use_cases_all}
\end{table}

First, we explain the advanced Scribble features, pointed in Table~\ref{table:use_cases_all}. 
In the table, the \textit{parameterised parallel composition} stands for the use of the construct \sCODE|par [i:1..N] {body}|, which is a shorthand for \sCODE{N} parallel \sCODE{body} branches, where \sCODE{i} is substituted with  values from \sCODE{1} to \sCODE{N} inside the \sCODE{body}. When the \sCODE{par} is followed by another block as in Listing~\ref{fig:big}, we call this block a join block (denoted as \textit{parallel with join} in Table~\ref{table:use_cases_all}). The sequence \sCODE|par {block 1} and {block 2} block 3| means that for  \sCODE{block3} to be executed, \sCODE{block1} and  \sCODE{blcok2} should be completed. For example, in Listing~\ref{fig:big} the message transfer \sCODE{done() from S to A; } can be completed only after both \sCODE{A} and \sCODE{B} have sent a message \sCODE{done} to \sCODE{S}. 

Another Scribble feature exploited in our examples is a \textit{subsession}. It is used to create modular Scribble code and is realised as a macro expansion. 
It is often combined with \textit{a group role} (Table~\ref{table:use_cases_all}, examples 7, 8, 11). A group role, for example, \sCODE{role Philosphers[1..N]},  specifies that there will be \sCODE{N} roles with the same type in the protocol. The roles share the same implementation, but at runtime there will be \sCODE{N} instances of that role and their behaviour might depend on initially assigned indices.

Next we present the examples. We first describe a global Scribble protocol for each use case based on 
specifications in \cite{Savina} and the implementations on github \cite{Savina_source}, and then provide
a couple of implementations in Python in Appendix~\ref{appendix:python:actors}. We also demonstrate faulty executions, that can be detected by our framework, on the Sleeping barber example in \fgr~\ref{fig:sleeping_barber}.
The use cases in Scribble and corresponding Python code can be found in \cite{OnlineAppendix}. 

Table~\ref{table:use_cases_all} shows a summary, specifying the
communication pattern and the MPST extension (if needed). For each
use case we give a brief description and the corresponding global Scribble protocol.
To express the use cases the previous version of Scribble has to 
be extended with (1) parametrised parallel composition and (2)
subsession with dynamic spawning of roles \cite{DBLP:conf/concur/DemangeonH12}.



{\lstset{caption={the Counting Protocol}, 
label=fig:counting_actor, boxpos=c,xleftmargin=0.2\textwidth, xleftmargin=0.2\textwidth}
\begin{SCRIBBLELISTING}
global protocol Counter (role P, role C)   	
{
	rec Loop { value(int) from P to C;}  	
	result(int) from C to P;}
}
\end{SCRIBBLELISTING}}

\paragraph{\bf Counting Actor.} This use case 
expresses an elaborated request-reply pattern, measuring the time
taken to send $n$ consecutive messages from a \sCODE{Producer} actor to a
\sCODE{Counter} actor. The accumulated value in the \sCODE{Counter}
is returned to the \sCODE{Producer}. From an actor perspective, the 
use case tests the mail box implementation which needs to store
the pending messages yet to be processed. Using session types
the communication interaction is succinctly specified
with recursion and point-to-point communication, expressing the
causalities between sending of a value (\sCODE{value(int)}) and the final reply (\sCODE{result(int)}).

{\lstset{caption={The Fibonacci Protocol}, 
label=fig:fib, boxpos=c,xleftmargin=0.2\textwidth, xleftmargin=0.2\textwidth}
\begin{SCRIBBLELISTING}[]
global protocol Fib(role P, role C) 
{
	request(int) from P to C;
	par { Fib(C as P, new C);}
	and { Fib(C as P, new C);}
	result(int) from C to P; }}
}
\end{SCRIBBLELISTING}}
\paragraph{\bf Fibonacci.} This use case involves the recursive creation of a predefined number of
actors using the naive recursive formula for computing fibonacci
numbers. The idea of the protocol is that an actor receives a request
to compute a value, then it spawns two additional actors and waits for
their response. Finally, it returns the sum of the responses to the
requester. The example tests the actor creation/destruction in the
actor implementation. The global Scribble protocol is given in Listing~\ref{fig:fib}. After receiving a
request (\sCODE{request}) from a \sCODE{Parent} actor, a \sCODE{Child} actor invokes the
same protocol type (\sCODE{Fib}) with a new \sCODE{Child} actor. To express the new
role creation, we use nested protocols (hereafter called
subprotocols), formally presented in \cite{DBLP:conf/concur/DemangeonH12}. A subprotocol
definition accepts either roles already active in the protocol (\sCODE|C as Parent|) 
or new roles of a known type (\sCODE|new C|). To express the
fork-join behaviour required by the example we use parallel
composition 
with continuation (the actions after a \sCODE{par} in a protocol body). Thus the specified protocol imposes causality between sending the response (\sCODE{result})
and the completion of the proceeding interactions, but
no causality between the subprotocols in the parallel
branches.  

\begin{multicols}{2}[\captionof{lstlisting}{The Big protocol}]
\lstset{label=fig:big}
\begin{SCRIBBLELISTING}
global protocol Big (role A, 
	role B,role Sink as S)
{
	par { ping() from A to B;
				pong() from B to A;
				done() from A to S;} 
	and { ping() from B to A;
				pong() from A to B;
				done() from B to S;}
	done() from S to A;
	done() from S to B;
}
\end{SCRIBBLELISTING}
\end{multicols}
\paragraph{\bf Big.} The use case presents a many-to-many message passing scenario. Several
actors are spawned, each of which sends a \sCODE{ping} message to another
random actor. Then the recipient actor responds with a \sCODE{pong} message. A \sCODE{Sink} actor
is notified (via \sCODE{done} message) about the completion of these interactions. When the \sCODE{Sink} receives all replies, it
terminates the other actors. In an actor implementation, the benchmark measures the effects on contention on a recipient's actor mailbox. In session types the use case requires
a join primitive, expressed in Scribble as a continuation after the
parallel branches, similarly to the Fibonacci
use case. 
Note that the round robin sending cannot be precisely expressed in
Scribble. We have simplified the use case, giving only the type for two
ping-pong actors (\sCODE{A} and \sCODE{B} in Listing~\ref{fig:big}) and one \sCODE{Sink} actor. The protocol can be extended
for more ping-pong actors if the configuration is fixed in
advance. 

{\lstset{label =fig:ring,caption=The ThreadRing protocol, xleftmargin=0.1\textwidth, xrightmargin=0.1\textwidth}
\begin{SCRIBBLELISTING}
global protocol Ring(role Worker[1..N])	
{	
	rec Loop {
		data(int) from Worker[i:1..N-1] to Worker[i+1];
		data(int) from Worker[N] to Worker[1];
	continue Loop;}
}
\end{SCRIBBLELISTING}}
 
\paragraph{\bf{Thread Ring.}} In the ThreadRing use case an integer token message is passed around a ring of \sCODE{N} connected actors. The token value is decremented each time an actor receives it. The interactions continue until the token becomes zero. From an actor perspective, the benchmark tests the raw efficiency of message sending since it removes all contention of resources except between the sending and receiving actors. To express the protocol for each \sCODE{N}-ring configuration we use an extension of Scribble, called Pabble \cite{Pabble}, where multiple participants can be grouped in the same role and indexed. Indeed, the same use case is already presented in \cite{Pabble}. In the protocol in Listing~\ref{fig:ring}, \sCODE{Worker[i:1..N-1]} declares iteration from \sCODE{1} to \sCODE{N-1} on the interaction (\sCODE{Data(int)  from Worker[i] to Worker[i+1]}) over the bound variable \sCODE{i}. 


%
{\lstset{label=fig:kfork,caption=The KFork protocol, xleftmargin=0.1\textwidth, xrightmargin=0.1\textwidth}
\begin{SCRIBBLELISTING}
global protocol KFork(role master as M, role Worker[1..K])
{
	par[i:1..K] {
		rec Loop {
			choice at M { 
				data(int) from M to Worker[i];} 
			or {
				end from M to Worker[i];}}
}
\end{SCRIBBLELISTING}}
\paragraph{\bf{KFork. }} The application first creates \sCODE{K} worker actors and then sends each actor a total of $n$ messages in a round robin manner. Actors terminate after processing all messages. The benchmark tests the message throughput in the implementation. The global multiparty protocol is expressed as a parallel composition of \sCODE{K} binary protocols with an identical communication pattern, recursive interaction between a \sCODE{Master} and a \sCODE{Worker}. For more concise specification, since all parallel branches describe an identical communication pattern, 
we leverage the parallel construct by introducing indexing on parallel branches. The syntax 
\sCODE|par[i:1..N] {body}| defines replication of \sCODE{body} for each value of \sCODE{i} between \sCODE{1} and \sCODE{K}. Listing~\ref{fig:fork_join_protocol} shows the global protocol for KFork using the new syntax.

{\lstset{caption=Refactored KFork Protocol, label=fig:fork_join_protocol}
\begin{SCRIBBLELISTING}
global protocol KFork(role Master as M, role Worker[1..K]) 
{ 
	par[i:1..K] { Sub(M, W[i]);}
} 
global protocol Sub(role M, role W)
{
	rec Loop {
		choice at M { 
			data(int) from M to W;} 
		or {
			end from M to W;}}
}
\end{SCRIBBLELISTING}}


\paragraph{\bf{Session patterns for concurrent group execution (Refactoring KFork).}}
We refactor the KFork global protocol, extracting the \sCODE{par} \sCODE{body} into a subprotocol. The global protocol with a group role is represented as a parallel composition of subprotocols. Indeed, this is an emerging pattern in all concurrent problems surveyed in the benchmark suit. The pattern appears in four use cases: Sleeping Barber, LogisticMap, BankTransaction and DiningPhilosophers.
\begin{figure}[tp]%
{\lstset{caption={Dining Philosophers, Sleeping Barber, Bank Transaction, Logistic Map}, 
label=fig:session_patterns, xleftmargin=0.1\textwidth, xrightmargin=0.1\textwidth}
\begin{SCRIBBLELISTING}
global protocol DiningPhilosophers(
	role arbitrator as A, role Philisopher[1..N])
{
	par[i:1..n] { DiningPhilosopher(A, Philisopher[i]); }
}
global protocol BankTransactions(
	role Teller as T, role Accounts[1..N])
{
	par[i:1..n-1] { BankTransaction(T, A[i], A[i+1]); }
}
global protocol LogisticMap(
	role Master as M, role Series[1..N], role Ratio[1..N])
{
	request from M to S[1..N];
	par[i:1..N] { SeriesMaster(S[i], R[i]);}
	response from S [1..N] to M;	
	stop from M to S[1..N];
}
\end{SCRIBBLELISTING}}
\end{figure}

A global protocol with a group role is split into smaller multiparty protocols - one for each role in the group role set. In addition, in LogisticMap and SleepingBarber a join multicast is required to signal the completion of the interaction and to end the protocol. In
Listing~\ref{fig:session_patterns} we give the four global protocols, where the recurring session pattern can be clearly observed. In the rest of the section we focus on the interactions between one of the roles from the group role set and give the protocols (DiningPhilosopher, SleepingBarberOneCustomer, BankTransaction, SeriesMaster respectively) which do not require the use of a group role. 

{\lstset{caption={The DiningPhilosophers protocol}, 
label=fig:dining_philosopher, xleftmargin=0.15\textwidth, xrightmargin=0.2\textwidth}
\begin{SCRIBBLELISTING}
global protocol DiningPhilosopher(
	role Arbitrator as A, role Philosopher as Ph)
{
	rec L {	choice at Ph {
						rec M {	
							req() from Ph to A;
								choice at A {
									yes() from A to Ph;
									continue L;} 
								or {
									no() from A to Ph;
									continue M;}}} 
					or { done() from Ph to A;}} 
}
\end{SCRIBBLELISTING}}

\paragraph{\bf The Dining Philosophers.} In this classic example, \sCODE{N} philosophers are competing to eat while sitting around a round table. There is one fork between each two philosophers and a philosopher needs two forks to eat. The challenge is to avoid deadlock, where no one can eat, because everyone is possessing exactly one fork. In Savina, the problem is implemented using \sCODE{Arbitrator} actor that coordinates the resources (the forks) between \sCODE{Philosopher} actors. The \sCODE{Arbitrator} either accepts (\sCODE{yes()}) a \sCODE{Philosopher} request or rejects (\sCODE{no()}) it, depending on the current resources allocated. If rejected, the \sCODE{Philosopher} keeps sending request (\sCODE{req}) until the request is accepted. The global protocol in Listing~\ref{fig:dining_philosopher} specifies the interactions between an \sCODE{Arbitrator} and one \sCODE{Philospher}, defining the causalities using recursion and choice. 

\newcommand{\CB}{\sCODE{B}}
\newcommand{\CR}{\sCODE{R}}
\newcommand{\CC}{\sCODE{C}}
\newcommand{\CS}{\sCODE{S}}

\paragraph{\bf{Sleeping Barber.}} The use case presents another classic synchorisation problem. A barber is waiting for customers. When there are no customers the barber sleeps. When a customer arrives, he wakes the barber or sits in one of the waiting chairs in the waiting room. If all chairs are occupied, the customer leaves. Customers repeatedly visit the barber shop until they get a haircut. The key element of the solution is to make sure that whenever a customer or a barber checks the state of the waiting room, they always see a valid state. The problem is implemented using a \sCODE{Selector} actor that decides which is the next customer, \sCODE{Customer} actors, a \sCODE{Barber} actor and a \sCODE{Room} actor. Note that, differently from the previous use cases, the protocol between the \sCODE{Customers} and the other actors is multiparty (not binary). To guarantee a valid synchronisation, precise message sequence should be followed. While the use cases so far demonstrated succinct recursion and request-reply patterns, interleaved together, this one demonstrates a communication structure with long sequence of interactions, where preserving the causalities is the key to achieving deadlock free communication. 

  The implementation is $\sim$200 LOC in Akka ($\sim$100 LOC in
  Python), where each of the four actors send messages to the other
  three actors. The implementation is error-prone, because the type of
  the actors and whether they can be safely compose is not obvious.
  Sending the wrong message type, sending to the wrong actor or not
  sending in the correct message sequence may lead to deadlocks,
  errors which initial cause is hard to be identified, or wrong
  computation results. Errors can even go unnoticed resulting in
  orphan actors. Here we give examples of several faulty executions
  and explain how they are prevented by our framework (for every error
  example, we highlight in \fgr~\ref{fig:sleeping_barber} a wrong
  implementation of a message transfer and the corresponding line in
  the global protocol):

\begin{description}
\item[Deadlock] (line~\ref{l:dl}): 
  After receiving message \sCODE{full} the Customer \sCODE{C} sends a
  message \sCODE{returned} to the Room \sCODE{R} instead of sending it
  to the Selector \sCODE{S}. Then the \sCODE{C} will wait to be
  reentered again, while \sCODE{S} will wait to receive
  \sCODE{returned} or \sCODE{done} from \sCODE{C} (according to
  Line~\ref{l:ut}), resulting in a deadlock. Moreover, the other
  actors are stuck as well (\sCODE{R} is waiting on \sCODE{S};
  \sCODE{B} is waiting on \sCODE{R}). Our framework will detect that the
  message \sCODE{returned} is sent to the wrong actor.

\item[Unexpected termination] (line~\ref{l:ut}):
  In this scenario, \sCODE{C} again sends a wrong message to \sCODE{R},
  but contrary to the above example, it sends a message that \sCODE{R}
  can handle (the message \sCODE{done}). If there is no verification at
  \sCODE{R} on the sender of the messages, \sCODE{R} will execute the
  \sCODE{done} handler and as a result will terminate (\sCODE{done}
  should be sent by \sCODE{S} to terminate the actor \sCODE{R} when all
  customers have been processed). When the Barber \sCODE{B} tries to
  send \sCODE{next} message to \sCODE{R} it will get a runtime exception
  that the actor \sCODE{R} does not exist. Such an exception can be
  very misleading because it points to a problem between the Barber
  and the Room, while the actual error is in the code for the
  Customer. Our framework will detect this inconsistency at the right (at role \sCODE{C})
  place therefore will not allow the error to be propagated, masking
  the real cause.

\item[Orphan actors] (line~\ref{l:oa}): 
  The actor \sCODE{R} sends a wrong message (\sCODE{don} instead of
  \sCODE{done}) to \sCODE{B} (or doesn't send a message at all). Therefore, \sCODE{B} will not terminate. However,
  everybody else will and so will the protocol, leaving the orhan
  actor \sCODE{B}. Our monitor will detect the wrong message that
  \sCODE{B} is sending.
\end{description}

\begin{figure}[tp]
\small
\begin{minipage}{0.5\linewidth}
\lstset{numbers=left}
\begin{SCRIBBLELISTING}
global protocol SleepingBarber(role barber as B, 
role selector as S, role room as R)
{rec L {
	  choice at S {
	  par {	  
	  S introduces C;}
	  enter() from S to R; 
	  choice at R {
			full() from R to C;
			`returned() from C to S;` *@\label{l:dl}@*
			continue L; }
		} or {
		wait() from R to C;
			 enter() from R to B;
			 start() from B to C;
			 stop() from B to C;
			 `done() from C to S;` *@\label{l:ut}@*
			 next() from B to R;
		}
		  wait() from R to B;
		  continue L;
	  } or {
	   	done() from S to R;
	    `done() from R to B;` *@\label{l:oa}@* }}
\end{SCRIBBLELISTING}
\end{minipage}
\begin{minipage}{0.45\linewidth}
\begin{PYTHONLISTING}



# Deadlock
class CustomerActor... 
  @role(c, S)	
	def full(self, c):	   
		# c.S.send.returned()
		`c.R.send.returned()` 



...
  @role(c, B)	
	def stop(self, c):	  
		# c.S.send.done() 
		`c.R.send.done()` 		

# Orphan Actors
class RoomActor... 
    @role(c, S)	
	def done(self, c):	 
		# c.B.send.done()  
		`c.B.send.don()` 
\end{PYTHONLISTING}
\end{minipage}
\caption{The Sleeping Barber protocol (left) and corresponding faulty implementations (right)}
\label{fig:sleeping_barber}
\end{figure}

\paragraph{\bf{Cigarette Smoker.}}
This problem involves \sCODE{N} smokers and one arbiter. The arbiter
puts resources on the table and smokers pick them. A smoker needs to
collect $k$ resources to start smoking. The challenge is to avoid
deadlock by disallowing a competition between smokers 
from picking up resources.
This is done by delegating the control to the arbiter, who decides (in
a random manner) which smoker to send the resource to. The session
type for the use case, given in Listing~\ref{fig:sig:smoker:actor}, combines round-robin pattern, sending a random
smoker a messageto smoke (\sCODE{StartSmoking}), with multicast,
iterating through all the smokers notifying them to exit (\sCODE{exit from A to S[i..n];}).

{\lstset{caption=The CigaretteSmoker protocol, label=fig:sig:smoker:actor}
\begin{SCRIBBLELISTING}
global protocol CigaretteSmoker(role Arbiter as A, 
	role Smoker[1..N] as S)
{
	rec Loop {
		choice at A {
			start_smoking() from A to S[i];
			started_smoking() from S[i] to A;
			continue Loop;
		} or {
			exit() from A to S[i..n]; }}}
\end{SCRIBBLELISTING}}


\paragraph{\bf{Bank Transaction.}} The communication pattern in this use case is a synchronous request-reply chain. Three actors (a \sCODE{Teller}, a source account \sCODE{SrcAcc} and a destination account \sCODE{DestAcc}) synchronise to perform a debit from one account to another. The \sCODE{Teller} is not allowed other operations until a confirmation that the transaction is complete is received from the source account. The two synchronous request-reply patterns are expressed as a message sequence in the BankTransaction protocol in Listing~\ref{lst:bank:transaction}. 

{\lstset{caption=The BankTransaction protocol, 
label=lst:bank:transaction,
xleftmargin=0.1\textwidth, xrightmargin=0.2\textwidth}
\begin{SCRIBBLELISTING}
global protocol BankTransaction(role Teller as T, 
	role SrcAcc as S, role DestAcc as D)
{   
		credit from T to S;
    debit from S to D;
    reply from D to S;
    reply from S to T; 
 }
\end{SCRIBBLELISTING}}

\paragraph{\bf{Logistic Map Series.}} This is another example of a synchronous request-reply pattern. The goal is to calculate a recurrence relation: 
$x_{n+1} = rx_n(1-x_n)$. The implementation requires an actor for a manager (\sCODE{Master}), a set of term actors (\sCODE{Series}) and a set of ratio actors (\sCODE{Ratio}). The \sCODE{Ratio} actors encapsulate the ratio \textit{r} and compute the next term $x_{n_1}$ from the current term $x_n$.
The \sCODE{Master} sends request to the \sCODE{Series} actor to compute a result. Then \sCODE{Series} actors delegate the calculation to a \sCODE{Ration} actor and wait for synchronous reply to update their value of $x$. Thus, the MasterSeries subprotocol is an instance of a request-reply pattern (from \sCODE{Series} as a requester to \sCODE{Ratio} role as a replier), which we have already shows and thus we omit.

\begin{multicols}{2}[\captionof{lstlisting}{The ProducerConsumer protocol}]
\lstset{label=fig:pcbounded}
\begin{SCRIBBLELISTING}
global protocol PCBounded(
	role Buffer as B, 
	role Producer as P, 
	role Consumer as C)
{
rec Loop {
	produce() from B to P;
	par {
		choice at P {
			dm() from P to B;
			continue Loop;
		} or {
			exit() from P to B;
		}
	} and {
		dm() from B to C;
		more() from C to B;}}
}
\end{SCRIBBLELISTING}
\end{multicols}

\paragraph{\bf{Producer Consumer with Bounded Buffer.}} The overall
scenario of this use case is that producers push work into 
buffers, while consumers pull from buffers. The implementation uses actors
for producers, consumers and a buffer. The \sCODE{Buffer} actor is
concurrently receiving messages from both \sCODE{Producer}s and
\sCODE{Consumer}s. Thus, the benchmark measures also the contention
overhead of the actor implementation. The Scribble protocol is given in Listing~\ref{fig:pcbounded}.  
To represent the pattern
(producers are sending messages concurrently, but there is causality
between the message send from a \sCODE{Producer} and the message send
to the \sCODE{Consumer}). We use a combination of
recursion and parallel branches. In each recursive execution, we
require \sCODE{Consumer}s and \sCODE{Producer}s to send or receive
concurrently. At the same time, the \sCODE{Producer} can keep sending
(starting a new recursive execution), before the \sCODE{Consumer} has
finished their
branch. 




\section{Related Work}
\label{sec:related}	
\paragraph{\bf Behavioural and session types for actors and objects}
There are several theoretical works that have studied the behavioural
types for verifying actors \cite{MostrousV11,Crafa}. The work 
\cite{Crafa} proposes a behavioural typing system for an actor
calculus where a type describes a sequence of inputs and outputs
performed by the actor body.  In \cite{MostrousV11}, a concurrent
fragment of Erlang is enriched with sessions and session
types. Messages are linked to a session via correlation sets (explicit
identifiers, contained in the message), and clients create the required
references and send them in the service invocation message. The
developed typing system guarantees that all within-session messages
have a chance of being received. The formalism is based on 
only binary sessions.  


Several recent papers have combined session types, as specification of
protocols on communicating channels, with object-oriented
paradigm. 
A work close to ours is \cite{GayVRGC10}, 
where a session channel is stored as a field of an
object, therefore channels can be accessed from different
methods. 
They explicitly specify the (partial) type for each method
bound to a channel. 
Our implementation also allows modularised sessions based on
channel-based communication.  
Their work  \cite{GayVRGC10} is mainly 
theoretical and gives a static typing system based on binary
session types. Our work aims to provide a design and implementation 
of runtime verification based on 
multiparty session types, and is 
integrated with existing development frameworks  
based on Celery \cite{celery} and AMQP \cite{AMQP}. 


The work in \cite{Caires08} formalises behaviours of non-uniform
active objects where the set of available methods may change
dynamically. It uses the approach based on spatial logic for a fine
grained access control of resources.
Method availability in \cite{Caires08} 
depends on the state of the object in a similar way as ours. 

See \cite{WG3} for more general comparisons between session 
types and other frameworks in object-oriented paradigms.  




\paragraph{\bf Other Actor frameworks}
The most popular actor's library (the AKKA framework \cite{AKKA} in Scala studied in
\cite{TasharofiDJ13}) supports FSM verification mechanism (through inheritance) and typed channels. Their channel typing is simple so that 
it cannot capture structures of communications such as sequencing,
branching or recursions. These structures ensured by session types are the key
element for guaranteeing deadlock freedom between multiple actors. 
In addition, in \cite{AKKA}, channels and FSMs are
unrelated and cannot be intermixed; on the other hand, in our approach, we rely on
external specifications based on the choreography (MPST) and the FSMs usage
is internalised (i.e.~FSMs are automatically generated from a global type),
therefore it does not affect program structures. 


Several works study an extension of actors in multicore control
flows. 
%
%
Multithreaded active objects \cite{MAO13} allow several threads to
co-exist inside an active object and provide an annotation system to
control their concurrent executions. 
Parallel Actor Monitors (PAM) \cite{PAM14} is another framework for
intra actor parallelism. PAMs are monitors attached to each actor that schedule the execution of the messages based on synchronisation constraints. 
Our framework also enables multiple control flows inside an actor as \cite{MAO13,PAM14}. In addition, we embed monitors inside actors in a similar way as \cite{PAM14,MAO13} embed schedulers. 
The main focus of the monitors in \cite{MAO13,PAM14} is scheduling the
order of the method executions in order to optimise the actor
performance on multi-core machines, while our approach aims to provide
explicit protocol specifications and verification among interactions
between distributed actors in multi-node environments.

The work \cite{JCoBox} proposes a technique to have multiple cooperatively scheduled tasks within a single active object similar to our notion of cooperative roles within an actor. 
The approach is implemented as a Java extension with an actor-like concurrency 
where communication is based on asynchronous method calls 
with objects as targets.
They resort to RMI for writing distributed implementations and do
not allow specifying sequencing of messages (protocols) like ours. 


The work \cite{ARC06} proposes a framework of three layers for actor 
roles and coordinators, which resembles roles and
protocol mailboxes in our setting. Their specifications focus on QoS
requirements, while our aim is to describe and ensure 
correct patterns of interactions (message passing).

Comparing with the above works, our aim is to provide effective
framework for multi-node environments 
where actors can be distributed transparently 
to different machines and/or cores.  

Several extensions of the actor model suggest more complex synchronization of the actor inbox such as multi-headed messages with guards \cite{Sulzmann:2008}, join-style actors \cite{JoinCalculus} and synchronous replies \cite{Varela:2001}. These works focus on more expressive pattern matching for precise and correct dispatch of the messages to the corresponding message handlers implemented in an actor. Thus, the synchronisation constraints are specified on the local actor behaviour. The emphasize on our work is to express the synchronization constraints globally, as part of the overall choreography of an actor system, and infer (through projection) the local behaviour of a single actor.



\section{Conclusion}
\label{sec:conclusion}
We propose an actor specification and verification framework based on
multiparty session types, providing a Python library with actor
communication primitives. 
Cooperative multitasking within sessions allows for combining
active and reactive behaviours in a simple and type-safe way.  
The framework is naturally integrated with 
Celery \cite{celery} which uses advanced message queue protocol 
(AMQP \cite{AMQP}), and  
 uses effectively its types for realising the key mechanisms  
such as the actor discovery. 
We demonstrated that the overhead of our implementation 
is very small. 
We then show that programming in session actors 
is straightforward 
by implementing and rewriting usecases from \cite{AKKA}. 
To our best knowledge, no other work links FSMs, actors and choreographies in
a single framework. As a future work, we plan to extend 
to other main stream 
actor-based languages such as Scala and Erlang to test the generality of our framework. 
Also a number of additional research items regarding the expressiveness of session types still need to be carried out. Among them: how round-robin communication (as found in the rest of the usecases in \cite{Savina}) can be addressed, and ...
As actor languages and frameworks are becoming more popular, we believe that
our work would offer an important step for writing correct 
large-scale actor-based communication programs.


\paragraph{\bf Acknowledgement}
We thank Ask Solem for his support and guidance. We also thank the reviewers for their valuable comments on the early version of this article. This work has been
partially sponsored by VMWare, Pivotal and EPSRC EP/K011715/1,
EP/K034413/1 and EP/L00058X/1, and EU project FP7-612985 (UpScale).




\footnotesize

\appendix
\label{appendix}
\section{Implementations}
\label{appendix:python:actors}
Here we present the corresponding Python code for some of the example. 

\begin{figure}[tp]%
{\lstset{caption={Fibonacci in Session Actors}, 
label=fig:fib_impl, xleftmargin=0.15\textwidth, xrightmargin=0.2\textwidth, 
numbers=left}
\begin{PYTHONLISTING}
@protocol(c, Fibonacci, P, Ch)
class Fibonacci(Actor):
	def __init__(self, first = False):
		self.ready = false
		self.current = 0		
		self.first = first
		
	@role(c, P)	
	def request(self, c, n):
			if n<=2:
				c.P.send.response(1)
			else: 
				fib1 = c.subprotocol("Fibonacci", 
					P=self, Ch=new Fibonacci())
				fib1.C.send.request(n-1)
										
				fib2 = c.subprotocol("Fibonacci", 
					P=self, Ch=new Fibonacci())
				fib2.C.send.request(n-2)

	@role(c, Ch)		
	def response(self, c, n):
		if self.ready:
			if self.is_first:
				print 'The result is', n + self.current 
			else: 
				c.P.send.response(n + self.current)
		else:
			self.ready = True
			self.current = n
\end{PYTHONLISTING}}
\end{figure}

We give in Listing~\ref{fig:fib_impl} the implementation of the Fibonacci protocol. The \pCODE{first} argument in the initialisation of a \pCODE{Fibonacci} actor specifies if this is the initial actor to receive the number to be computed. 
As prescribed by the protocol, two message handlers, \pCODE{request} and \pCODE{response}, are implemented. The latter is the message received by the parent role \pCODE{P}, while the former is received from the child role \pCODE{Ch}. Line 13--17 illustrate a new method on the \pCODE{ActorRole} class for creating subprotocols. Note that the \pCODE{response} method should be called twice (once for each subprotocol). When the first response message is received the payload is saved temporary in the actor field \pCODE{self.current} (line 30). When the second response is received, its value is added to \pCODE{self.current} and the sum is returned to the parent actor (line 27). 

\begin{figure}[tp]%
{\lstset{caption={Counter in Session Actors}, 
label=fig:counter:impl, xleftmargin=0.15\textwidth, xrightmargin=0.2\textwidth, 
numbers=left}
\begin{PYTHONLISTING}
@protocol(cc, Counter, C, P)
class Counter(Actor):
	def __init__(self):
		self.current = 0
	
	@role(cc)	
	def val(self, n):
			self.current+=n
			
	@role(cc, P)	
	def retrieve_message(self, cc):
		cc.P.send.res(self.current)

@protocol(cc, CounterProtocol, C, P)
class Producer(Actor):
	@role(cc, self)	
	def join(n, cc):
		for i in range(1, n):	
			cc.C.send.val(1)	
		cc.C.send.retrieve_message()
			
	@role(cc, C)		
	def res(self, n)
		print 'The result is', n
\end{PYTHONLISTING}} 
\end{figure}

\paragraph{\bf Counter}
We give in Listing~\ref{fig:counter:impl} the implementation of the Counter protocol. We have two Actor classes, \pCODE{Counter} and \pCODE{Producer}. 
The interactions start with the \pCODE{join} method of the \pCODE{Producer} class. The body defines consecutive sending of $n$ messages (invoking the message handler \pCODE{val}) of the \pCODE{Counter} actor (role \pCODE{C} in the protocol CounterProtocol). In the \pCODE{val} method the counter actor accumulates the number of received messages as part of its internal state (\pCODE{self.current}). After the iteration is completed, the \pCODE{Producer} sends \pCODE{retrieve_message} to the \pCODE{Counter}, asking for the result (line 20). The annotation on \pCODE{retrieve_message} specifies that the receiver of the message is the role instance \pCODE{cc} and the sender is the role \pCODE{P} (Producer). In the body of \pCODE{retrieve_message}, the \pCODE{Counter} sends to \pCODE{P} (\pCODE{cc.P.res}) the final result (\pCODE{self.current}).

\begin{figure}[tp]%
{\lstset{caption={DiningPhilosphers in Session Actors}, 
label=fig:dining:impl, xleftmargin=0.15\textwidth, xrightmargin=0.2\textwidth, 
numbers=left}
\begin{PYTHONLISTING}
@protocol(c, DiningPhilosophers, Ph, A)
class Philosopher(Actor):
	def __init__(self, no):
		self.no = no	
		
	def join(self, c):
		c.A.send.req(self.no)
		
	def yes(self):
		self.become(self.eat)		
		
	@role(c, self)
	def eat(self):
		if full:
			c.A.send.done()

	def no(self):
		c.A.send.req(self.no)
			
@protocol(c, DiningPhilosophers, A, Ph)
class Arbitrator(Actor):
	
	@role(c, Ph)	
	def req(self, c, sender=ph_no)	
		if self.free_forks():
			c.Ph[ph_no].send.yes()
		else:		
			c.Ph[ph_no].send.no()
	@role(c, Ph)	
	def done(self, c):
		print 'The dinner is over'
		
\end{PYTHONLISTING}}
\end{figure}

\paragraph{\bf DiningPhilosopher}
We give in Listing~\ref{fig:dining:impl} the implementation of the DiningPhilosophers protocol. For sake of space and clarity, the implementation shows only the interactions for the \pCODE{Arbitrer} and \pCODE{Philospher}, abstracting from the resource management done by the \pCODE{Arbitrer} (in \pCODE{free_forks()}). For a \pCODE{Philospher} the protocol starts from the method \pCODE{join}. Using its role instance, the actor does the first request to the \pCODE{Arbitrer} to eat (\pCODE{c.A.req(self.no)}). To distinguish between \pCODE{Philosphers}, each one is given a number when initialised and when a \pCODE{Philosopher} sends a message to an \pCODE{Arbitrer} (\pCODE{c.A.req}) it also sends its number. The \pCODE{Arbitrer} uses this information to identify \pCODE{Philosopher} that is waiting for the reply(\pCODE{c.Ph[ph_no]}). Note that in line 26 and in line 28 the \pCODE{Arbitrer} does not know which \pCODE{Philospher} is to receive a message (the annotation only specifies that the sender is a role of type \pCODE{Ph} (\pCODE{Philosopher})). When a \pCODE{Philosopher} receives a \pCODE{yes} reply it sends a message to itself to eat (\pCODE{self.become(self.eat)}). If \pCODE{no} is returned (line 17), the \pCODE{Philosopher} repeats its request to the \pCODE{Arbitrer} (\pCODE{c.A.req}). 

\begin{figure}[tp]%
{\lstset{caption={CigaretteSmoker in Session Actors}, 
label=fig:sleeping:impl, xleftmargin=0.15\textwidth, xrightmargin=0.2\textwidth, 
numbers=left}
\begin{PYTHONLISTING}
@protocol(c, CigaretteSmoker, A, S)
class Arbiter(Actor):
	def __init__(self, n):
		self.n = n
		 
	def join(self, c):
		# sends a message to a random smokers		
		radom = generate_random()
		c.S[random].send.start_smoking()
		
	@role(c, S)
	def started_smoking(self):
		if enough:
			for i in range(1, n):
				c.S[i].send.exit()
		else: 
			radom = geenrate_random()
			c.S[random].send.start_smoking()	
	
@protocol(c, CigaretteSmoker, S, A)
class Smoker(Actor):	
	@role(c, A)	
	def start_smoking(self):
		c.A.send.started_smoking()			
			
	@role(c, A)
	def exit(self):
		print 'I am done.' 
\end{PYTHONLISTING}}
\end{figure}

\paragraph{\bf CigaretteSmoker}
We give in Listing~\ref{fig:sleeping:impl} the implementation of the CigaretteSmoker protocol. Similar to the DiningPhilospher, the protocol is initialised by giving the number of participating of Smokers, e.g $n$. The \pCODE{Arbitrer} initiates the first interaction through its \pCODE{join} method, where it sends a message \pCODE{start_smoke} to a random smoker (it chooses random smoker from the list of smokers-- \pCODE{S[random]}). 
The \pCODE{Smoker} actor has a handler for \pCODE{start_smoke}. The handler is annotated with a \pCODE{@role}, specifying this is a handler for the role smoker \pCODE{c} and the caller for that handler will be the role \pCODE{A}. Then the \pCODE{Smoker} notifies the \pCODE{Arbitrer} so the \pCODE{Arbitrer} can proceed and give resources to other \pCODE{Smokers}. When the \pCODE{Arbitrer} decides to end the smoking it sends \pCODE{exit} message to everyone.   

\newpage\phantom{blah}\newpage\phantom{blah}
\relax

\end{document}